\documentclass[]{aa}  
\usepackage{array, multirow, graphicx}
\usepackage[dvipsnames]{xcolor}
\usepackage{wasysym}[integrals]
\usepackage{amssymb}
\usepackage{pifont}
\usepackage{siunitx}
\usepackage{booktabs, tabularx}

\usepackage{txfonts}
\usepackage[]{hyperref}
\hypersetup{colorlinks,citecolor=blue, linkcolor=blue, urlcolor=blue, breaklinks=true}
\usepackage{cellspace, makecell} 
\usepackage{mathtools}

\newcounter{magicrownumbers}

\newcommand{\tonda}[1]{\!\left ( #1 \right )}
\newcommand{\quadra}[1]{\left [ #1 \right ]}

\newcommand{\abs}[1]{\left | #1 \right |}

\newcommand{\eq}[1]{\begin{equation} #1 \end{equation}}

\newcommand{\RNum}[1]{\uppercase\expandafter{\romannumeral #1\relax}}

\newcommand{\cii}{[C\RNum{2}]$_{\rm 158\,\mu m}$}
\newcommand{\ci}[1]{[C\RNum{1}]$_{\rm #1\,\mu m}$}

\begin{document}

   \title{ALMA survey of a massive node of the Cosmic Web at $z\sim 3$}
      \subtitle{I. Discovery of a large overdensity of CO emitters}

   \author{A.~Pensabene
          \inst{\ref{unimib}}
         	\and
	S. Cantalupo
	\inst{\ref{unimib}}
		\and
	C. Cicone
	\inst{\ref{ITA}}
		\and
         R.~Decarli
         \inst{\ref{inaf-bo}}
         	\and
         M.~Galbiati
	\inst{\ref{unimib}}
		\and        
	M.~Ginolfi
	\inst{\ref{unifi-arcetri}}
		\and
	S.~de~Beer
	\inst{\ref{unimib}}
		\and	
	M.~Fossati
	\inst{\ref{unimib},\ref{inaf-brera}}
		\and
	M.~Fumagalli
	\inst{\ref{unimib},\ref{inaf-trieste}}
		\and
	T.~Lazeyras
	\inst{\ref{unimib}}
		\and	
	G. Pezzulli
	\inst{\ref{kapt}}
		\and
	A.~Travascio
	\inst{\ref{unimib}}
		\and
	W.~Wang
	\inst{\ref{unimib}}
		\and	
	J. Matthee
	\inst{\ref{ist-austria}}
		\and
	M. V. Maseda
	\inst{\ref{wisc}}
          }

   \institute{Dipartimento di Fisica ``G. Occhialini'', Universit\`a degli Studi di Milano-Bicocca, Piazza della Scienza 3, I-20126, Milano, Italy\\\email{antonio.pensabene@unimib.it}\label{unimib}
   		\and
Institute of Theoretical Astrophysics, University of Oslo, P.O. Box 1029, Blindern 0315, Oslo, Norway\label{ITA}
		\and
   	INAF-Osservatorio di Astrofisica e Scienza dello Spazio, Via Gobetti 93/3, I-40129 Bologna, Italy\label{inaf-bo}
		\and
Dipartimento di Astronomia e Scienza dello Spazio, Universi\`a degli Studi di Firenze, Largo E. Fermi 2, 50125 Firenze, Italy\label{unifi-arcetri}
		\and
	INAF-Osservatorio Astronomico di Brera, Via Brera 28, I-21021 Milano, Italy\label{inaf-brera}
		\and
	INAF-Osservatorio Astronomico di Trieste, Via G. B. Tiepolo 11, I-34143 Trieste, Italy	\label{inaf-trieste}
		\and
	Kapteyn Astronomical Institute, University of Groningen, Landleven 12, NL-9747 AD Groningen, the Netherlands\label{kapt}
		\and
	Institute of Science and Technology Austria (IST Austria), Am Campus 1, 3400 Klosterneuburg, Austria\label{ist-austria}
		\and
	Department of Astronomy, University of Wisconsin-Madison, 475 N. Charter St., Madison, WI 53706 USA\label{wisc}
             }
   \date{Received XXX accepted YYY}

 
\abstract{Sub-mm surveys toward overdense regions in the early Universe are essential to uncover the obscured star-formation and the cold gas content of assembling galaxies within massive dark matter halos. 
In this work, we present deep ALMA mosaic observations covering an area of $\sim 2\arcmin\times2\arcmin$ around MQN01 (MUSE Quasar Nebula 01), one of the largest and brightest Ly-$\alpha$ emitting nebulae discovered thus far surrounding a radio-quiet quasar at $z\simeq3.25$. Our observations target the 1.2- and the 3-mm dust continuum, as well as the carbon monoxide CO(4--3) transition in galaxies in the vicinity of the quasar. We identify a robust sample of eleven CO line-emitting galaxies (including a closely-separated quasar companion) which lie within $\pm 4000\,{\rm km\,s^{-1}}$ relatively to the quasar systemic redshift. A fraction of these objects are missed in previous deep rest-frame optical/UV surveys thus highlighting the critical role of (sub-)mm imaging. We also detect a total of eleven sources revealed in their 1.2-mm dust continuum with six of them having either high-fidelity spectroscopic redshift information from rest-frame UV metal absorptions, or CO line which place them in the same narrow redshift range. A comparison of the CO luminosity function (LF) and 1.2-mm number count density with that of the general fields points to a galaxy overdensity of $\delta > 10$. We find evidence of a systematic flattening at the bright-end of the CO LF with respect to the trend measured in blank fields. Our findings reveal that galaxies in dense regions at $z\sim3$ are more massive and significantly richer in molecular gas than galaxies in fields, hence enabling a faster and accelerated assembly. This is the first of a series of studies to characterize one of the densest regions of the Universe found so far at $z > 3$.}


   \keywords{galaxies: evolution --
   		    galaxies: high-redshift --
		    galaxies: ISM --
		    cosmology: large-scale structure of Universe --
		    submillimeter: galaxies
                    }
               
   \titlerunning{The ALMA view of MQN01 field}
   \authorrunning{Pensabene et al.}
   \maketitle
   
%

\section{Introduction}
\label{sect:introduction}
The formation and evolution of galaxies and active galactic nuclei (AGN) is believed to occur within a network of diffuse intergalactic gas (IGM) distributed along filaments and sheets on Mpc-scale structures \citep[the so-called ``Cosmic Web'';][]{Bond+1996}. Cosmological simulations predict that galaxy formation takes place in the densest regions of this structure where the assembly of galaxies is regulated by the complex interplay between accretion of gas from the Cosmic Web and gas ejection from galaxies into the IGM triggered by feedback mechanisms acting on galactic scales \citep[such as, star-formation and AGN-driven outflows, see, e.g.,][]{McNamara+2012, Pike+2014, Wilkinson+2018}. However, the details of these processes are still poorly understood. The collection of many lines of evidence over the years led to a growing consensus concerning the key role of the large-scale environment in shaping galaxies during their evolution, at least at $z\lesssim 1$. Elliptical galaxies are preferably found in clusters \citep[]{Oemler+1974, Davis+1976, Dressler+1980, Postman+1984, Dressler+1997, Boselli+2006} and tend to be passive (with red colors) while galaxies in fields are predominantly blue spirals exhibiting substantial star-formation activity \citep[see, e.g.,][]{Lewis+2002, Gomez+2003, Balogh+2004, Hogg+2004, Kauffmann+2004, Tanaka+2004, Park+2007, Peng+2010}. The situation at higher redshifts is however unclear. While on the one hand, some results suggest that the aforementioned trend holds up to $z\sim 2$ \citep[see, e.g.][]{Postman+2005, Muzzin+2012,Quadri+2012, Darvish+2016, Fossati+2017, Kawinwanichakij+2017, Perez-Martinez+2023, Taylor+2023}, others support the evidence of enhanced star formation activity in protoclusters at $z>2$ compared to the field galaxies (e.g., \citealt{Elbaz+2007, Cooper+2008, Ideue+2009, Tran+2010, Shimakawa+2018, Ito+2020}, but see, also, \citealt{Dominguez-Gomez+2023}), which is consistent with the hypothesis that galaxies in dense environments assemble their mass more rapidly and at earlier times \citep[see,][for a recent review and further discussion]{Alberts+2022}. In order to get further insights on how galaxies form and evolve in connection with their large-scale environment, it is therefore crucial to obtain a comprehensive view of overdense regions of galaxies at early epochs, especially during the peak of galaxy assembly and AGN activity at $z\sim 2-3$ \citep{Madau+2014}.

At high redshifts, there is an increasing contribution from dusty star-forming galaxies to the cosmic star-formation rate density \citep[see, e.g.,][]{Casey+2014, Dunlop+2017, Hodge+2021} meaning that a larger fraction of galaxies might remain undetected even in deep rest-frame optical/UV surveys, and possibly even in near-infrared (NIR) observations \citep[e.g.][]{Williams+2019, Yamaguchi+2019, Smail+2021, Smail+2023, Manning+2022}. Such galaxies can however be uncovered in the rest-frame far-infrared (FIR) band where the thermal emission by dust grains dominates the galaxy spectral energy distribution (SED). At such wavelengths, atomic fine-structure and molecular emission lines are the major coolants of the gas-phase galaxy interstellar medium (ISM), and they can therefore be targeted to trace the cold gas in galaxies \citep[see,][for a review]{CarilliWalter2013}.

With the advent of sensitive facilities in the (sub-)mm such as the Atacama Large (sub-)mm Array (ALMA), astronomers have started to map dense galaxy environments such as (proto-)clusters of galaxies at increasingly high redshift to understand how cold gas - the ultimate fuel of star-formation - and dust - which is a proxy of the galaxy metal enrichment - are affected in galaxy-rich environments with respect to galaxies living in isolation. Such studies have been mainly conducted by targeting carbon monoxide (CO) rotational lines, the singly-ionized atomic carbon transition \cii{}, as well as the FIR dust continuum at $z\sim 1-2$ \citep[e.g.][]{ Hayashi+2017, Noble+2017, Noble+2019, Rudnik+2017, Stach+2017, Coogan+2018, Williams+2022}, $z\sim 2-3$ \citep[e.g.,][]{WangT+2016, WangT+2018, Lee+2017, Castigliani+2019, Gomez-Guijarro+2019, Tadaki+2019b, Champagne+2021, Jin+2021, Aoyama+2022}, up to z $\sim 3-4$  \citep[see, e.g.,][]{Hodge+2013, Miller+2018, Oteo+2018, Umehata+2019, Hill+2020, Polletta+2022}. These investigations led to the discovery of numerous gas-rich galaxies in (the core of) galaxy (proto-)clusters. Nevertheless, despite all these efforts, the emerging picture is still unclear and contradictory with tentative evidence of an enhanced star-formation rate, gas and dust fraction, and molecular gas excitation in clustered galaxies, at least at $1<z<2$.

Crucially, the aforementioned works highlight the importance of sub-mm (pseudo-)blind surveys toward galaxy dense regions at high-$z$ to probe galaxy CO luminosity functions (LFs) and the spectral-line energy distribution, as well as the mm number counts which can provide us with key clues on the physical processes that are acting in the nodes of the Cosmic Web. In this work, we present ALMA observations toward the MUSE Quasar Nebula 01 (MQN01) field. This field hosts a giant Ly-$\alpha$ emitting nebula initially discovered via a blind survey of bright radio-quiet quasars (or QSOs) at $3<z<4$ \citep{Borisova+2016} using the Multi Unit Spectroscopic Explorer (MUSE) mounted on the Very Large Telescope (VLT). The MQN01 nebula surrounding the QSO CTS G18.013 at $z=3.25$ is one of the largest nebulae ($\sim 30\arcsec$ corresponding to $\sim 230$ physical kpc) discovered in this survey that also exhibits a filamentary morphology. The largest Ly-$\alpha$ nebulae discovered so far are often associated with an overdensity of AGN and massive (dusty) star-forming galaxies \cite[see, e.g., ][]{Hennawi+2015, Cai+2017, Cantalupo+2017, Arrigoni-Battaia+2018b, Arrigoni-Battaia+2018, Umehata+2019, Nowotka+2022}. MUSE follow-up observations extending both the covered area and sensitivity limit mapped a large $\sim 4\,{\rm arcmin}^{2}$ area around MQN01, revealing a high concentration of Lyman Break Galaxies (LBGs) embedded in an extended Ly-$\alpha$ emitting structure (Cantalupo et al., in prep., Galbiati et al., in prep.). To obtain a full picture of this exceptional field we used ALMA to perform mosaic observations targeting the mm dust continuum and the CO(4--3) rotational transition in galaxies embedded in MQN01. This work is part of an extensive multiwavelength survey of the MQN01 field which has been conducted in the FIR to the X-rays regime using multiple facilities. Here, we report galaxy detections and field statistics obtained via our mm observations using ALMA. The full characterization of individual sources and their correlation with the Ly-$\alpha$ emitting gas will be presented in future works. 

This paper is structured as follow: in Sect.~\ref{sect:observations} we present our survey design, the acquired observations, the reduction of the data, and the ancillary datasets. In Sect.~\ref{sect:source_search}, we discuss the source extraction and the measurements of continuum fluxes and line luminosities of the selected candidates. In Sect.~\ref{sect:results}, we present our results which include the analysis of the CO LF, and the mm-continuum source number count density. We dedicate Sect.~\ref{sect:discussion} to the interpretation and discussion of our results and we compare them with previous works putting our findings into a more general context. Finally, in Sect.~\ref{sect:summary_conclusions} we draw our conclusions.

Throughout this paper we assume a standard $\Lambda\rm{CDM}$ cosmology with $H_0=67.7\,\si{km\,s^{-1}Mpc^{-1}}$, $\Omega_{m}=0.310$, $\Omega_{\Lambda}=1-\Omega_{m}$ from \citet{PlanckColl+2020}.

\begin{table*}[!htbp]
\caption{Characteristics of ALMA observations toward the MQN01 field.}  
\label{tbl:obs_summary}      
\centering   
\resizebox{\hsize}{!}{
\begin{tabular}{l c c}     
\toprule\toprule
										&	{\bf Band 3}										&{\bf Band 6}\\
\cmidrule(lr){1-3}
Tuning Central Wavelength\,$^{(1)}$					&	$2.94\,{\rm mm}$									&$1.26\,{\rm mm}$\\
SPW Central Frequencies						&	$95.00,\,96.80,107.20,109.00\,{\rm GHz}$					&$229.20,\,231.00,243.20,245.00\,{\rm GHz}$\\
SPW Bandwidth							&	$1.875\,{\rm GHz}$									&$1.875\,{\rm GHz}$\\		
\cmidrule(lr){1-3}
R.A. extrema (ICRS)							& (00:41:23.06, 00:41:41.49)								&(00:41:25.03, 00:41:39.51)\\
DEC. extrema (ICRS)						& (-49:37:55.58, -49:34:43.58)								&(-49:37:33.84, -49:35:07.14)\\
No. of Pointings							&	$27$												&$114$\\
HPBW Primary Beam (FOV)\,$^{(2)}$			& 	$53\rlap{.}{\arcsec}4$										&$23\rlap{.}{\arcsec}8$\\
Mosaic Spacing							&	$27\rlap{.}{\arcsec}4$										&$12\rlap{.}{\arcsec}1$\\
Sky Area Coverage\,$^{(3)}$					&	$6.1308\,{\rm arcmin^{2}}$							&$4.4051\,{\rm arcmin^{2}}$\\
\cmidrule(lr){1-3}
Number of Antennas	($12\,{\rm m}$)				&	$41-46$											&$40-44$\\
Baselines	(m)								&	$14.9-976.6$										&$14.6-500.2$\\
Synthesized Beam size\,$^{(4)}$				&  	${\rm 1\rlap{.}{\arcsec}41\times1\rlap{.}{\arcsec}29}$			&${\rm 1\rlap{.}{\arcsec}23\times1\rlap{.}{\arcsec}05}$\\
Synthesized Beam PA\,$^{(4)}$   				&  	${\rm -57.25\,{\rm deg}}$								&${\rm 89.92\,{\rm deg}}$\\
\cmidrule(lr){1-3}
Observation Dates							& 2022 January 6 $-$ May 19								&2022 April 3 $-$ 13\\
Flux Calibrator								&J2357-5311											&J2258-2758\\
Mean PWV (mm)\,$^{(5)}$					&$1.8-6.2$											&$0.3-2.4$\\
Total Time on Science Target 					& 	$16.3\,{\rm hours}$									&$3.3\,{\rm hours}$\\
RMS Representative Bandwidth\,$^{(6)}$			&$0.12\,{\rm  mJy\,beam^{-1}}$								&$42\,{\rm \mu Jy\,beam^{-1}}$\\
\bottomrule
\end{tabular}
}
\tablefoot{$^{(1)}$Central wavelength of the entire frequency setting. $^{(2)}$Half Power Beam Width (i.e., $\sim 1.13\times\lambda/D$, where $\lambda$ is the observed wavelength, and $D$ is the ALMA antenna diameter, see \citealt{Remijan+2019}) at the representative frequency of $109.00\,{\rm GHz}$ (band 3) and $245.00\,{\rm GHz}$ ({\rm band 6}). $^{(3)}$Observed sky angular area with primary beam sensitivity $\ge 50\%$. $^{(4)}$Synthesized beam size and position angle (PA) at the representative frequency in case of a {\rm natural} weighting scheme of the visibilities. $^{(5)}$Mean Precipitable Water Vapor during observations. $^{(6)}$The representative bandwidth is $100\,{\rm km\,s^{-1}}$ (corresponding to $36.358\,{\rm MHz}$) at the representative frequency of band 3, and $6.89\,{\rm GHz}$ (aggregate continuum) for band 6 observations.}
\end{table*}

\section{Observations and data processing}\label{sect:observations}
\subsection{Survey design}\label{ssect:survey_design}
We observed the MQN01 field with ALMA 12-m array using band 3 and 6 in Cycle 8 (Program ID. 2021.1.00793.S, PI: S. Cantalupo). The observations were designed to cover the entire field of view (FoV) of the MUSE mosaic ($\sim 4\,{\rm arcmin^{2}}$, corresponding to $\simeq16\,{\rm cMpc^{2}}$ at $z=3.25$) by performing a Nyquist-sampled mosaics following the standard hexagonal pattern to achieve a uniform sensitivity across the entire field. 

The band 3 mosaic consists of {\rm 27} pointings each with a Half Power Beam Width (HPBW) of $\simeq 53\arcsec$ resulting in a covered rectangular sky area of $\simeq 3\rlap{.}{\arcmin}0\times 3\rlap{.}{\arcmin}2$. Observations were carried out in the Frequency Division Mode (FDM). The frequency setup consists of four $1.875\,{\rm GHz}$-wide spectral windows (SPWs). We tuned two adjacent SPWs in the Upper Side Band (USB) centered respectively at $107.20\,{\rm GHz}$ and $109.00\,{\rm GHz}$ such that they encompass the CO(4--3) transition (rest-frame frequency $\nu_{\rm rest}=461.041\,{\rm GHz}$), as well as the underlying 3-mm dust continuum in a contiguous redshift bin of $\Delta z \simeq 0.15$ corresponding to $\Delta\varv = (-4000, +6100)\,{\rm km\,s^{-1}}$ around $z=3.25$. We tuned the other two SPWs in the Lower Side Band (LSB) covering a continuous bandwidth in the frequency range $94.06 - 97.74\,{\rm GHz}$. {\rm The total effective bandwidth of ALMA band 3 observations is $7.354\,{\rm GHz}$}. The native spectral resolution of the acquired data is $1.95\,{\rm MHz}$ ($\sim 5.4 \,{\rm km\,s^{-1}}$). Observations were carried out in eighteen Execution Blocks (EBs) during the period 2022 January 16 -- May 19 employing a total on source exposure time of $16.4\,{\rm hours}$ and maximum antenna baseline of $976.6\,{\rm m}$. During the executions, the quasars J0025-4803 and J2357-5311 were observed as phase and flux calibrator, respectively.

The band 6 mosaic consists of 114 pointing with ${\rm HPWB \simeq 24\arcsec}$ covering a sky area of $\simeq 2\rlap{.}{\arcmin}3\times 2\rlap{.}{\arcmin}4$. We employed a frequency setup in the FDM with two pairs of adjacent $1.875\,{\rm GHz}$-wide SPWs that we disposed to cover the 1.2-mm dust continuum together with the CO(9--8) ($\nu_{\rm rest}=1036.912\,{\rm GHz}$), as well as the adjacent transitions of the Hydroxyl Ion ${\rm OH^{+}}$($1_{1}$--$0_{1}$) ($\nu_{\rm rest}=1033.119\,{\rm GHz}$) that are redshifted in the ALMA band 6 at $z\simeq3.25$. The two SPWs in the USB are centered respectively at $243.20\,{\rm GHz}$, and $245.00\,{\rm GHz}$ sampling a contiguous CO(9--8) line redshift bin of $\Delta z\simeq0.06$, corresponding to $\Delta\varv = (-2400, +2100)\,{\rm km\,s^{-1}}$ around $z=3.25$. The SPWs in the LSB cover the frequency range $228.26 - 231.94\,{\rm GHz}$. The effective total bandwidth is $7.35\,{\rm GHz}$ with a native frequency sampling of $3.9\,{\rm MHz}$ ($\sim 4.8\,{\rm km\,s^{-1}})$. The observations were carried out in four EBs during the period 2022 April 3 - 13 employing a total on source observation time of $3.3\,{\rm hours}$ and a maximum baseline of $500.2\,{\rm m}$. For such observations, the quasars J0025-4803 and J2258-2758 were used as phase and flux calibrator, respectively.

\begin{figure*}[!t]
   	\centering
   	\resizebox{\hsize}{!}{
      		\includegraphics{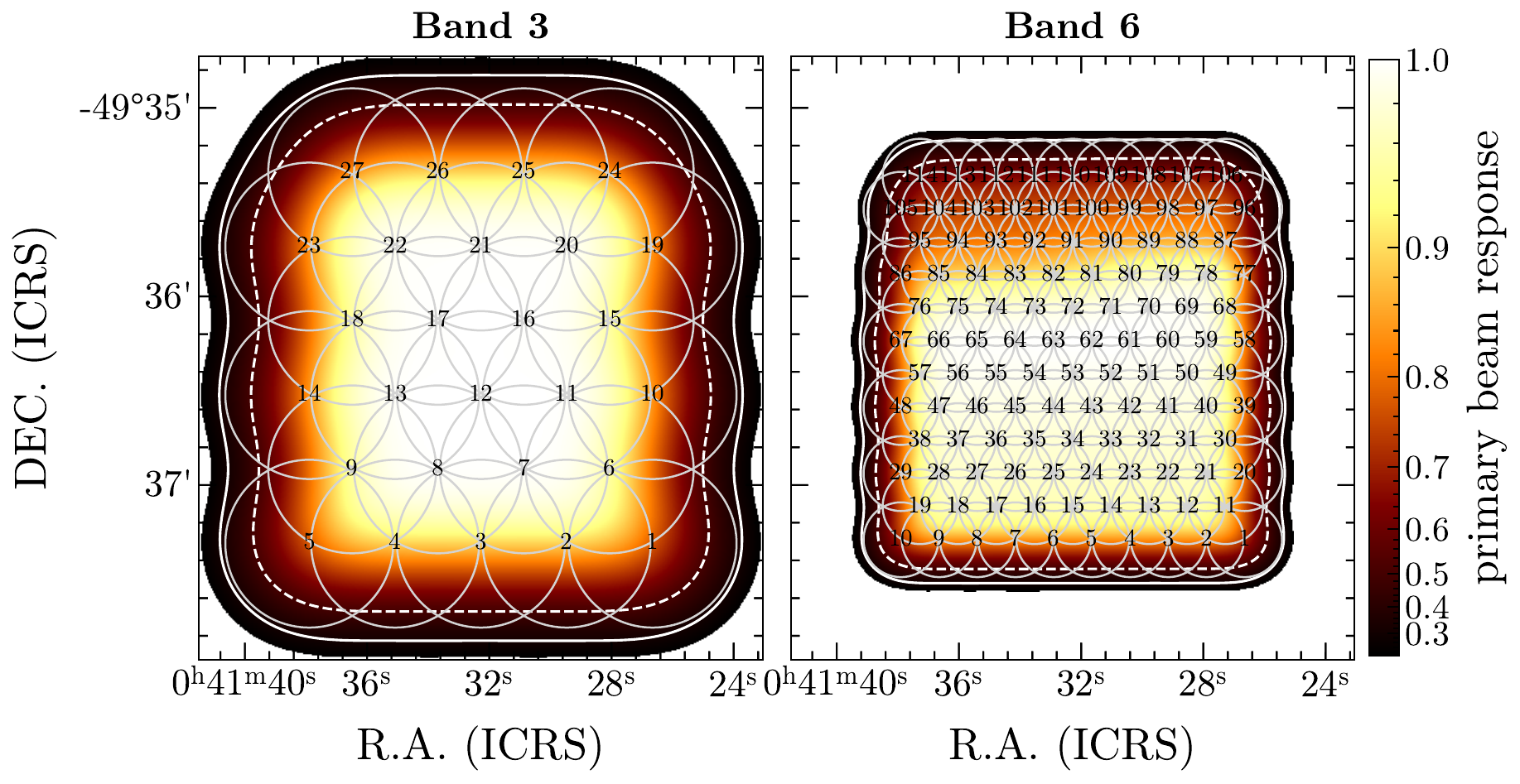}}
       \caption{Combined primary beam response for our {\rm ALMA mosaics at the representative frequency of $109.0 \,{\rm GHz}$ (band 3, {\it left panel}) and for the aggregate continuum mosaic image of ALMA band 6 ({\it right panel})}. The white contours correspond to a primary beam response of $30\%$ (solid line) and $50\%$ (dashed line). The circles show the disposition of the pointings with diameter equal to the Half Power Beam Width (HPBW) of the ALMA 12m antennas at the reference frequency of the setup.}
       \label{fig:mos_pointings}
   \end{figure*}

In Table~\ref{tbl:obs_summary} we report the details of the observations presented in this work. In Fig.~\ref{fig:mos_pointings} we show the combined primary beam (PB) response of the mosaics in the two different bands together with the disposition of the pointings. The last EB of band 6 observations has been affected by increased noise during the last two scans, impacting the sensitivity of the mosaic pointings no. 77-114. As a result, the sensitivity in the Northern part of the FoV is about $30\%$ lower relatively to the mosaic center.

\subsection{Data reduction}\label{ssect:data_reduction}
We performed data reduction using the Common Astronomy Sofware Application \citep[{\sc CASA};][]{McMullin+2007, Hunter+2023}. We calibrated the data of both band 3 and 6 observations by running the pipeline scripts ({\tt scriptForPI}) delivered alongside with the raw Measurement Sets (MSs) by using the CASA pipeline version 6.2.1. In the case of band 6 data, the calibration includes the renormalization correction related to the ALMA amplitude normalization strategy. We imaged the ALMA band 3 and 6 visibilities by using the CASA task {\tt tclean} by adopting a natural weighting scheme, and by using ``mosaic'' as a gridding convolution function. We set the phase center at the coordinates ICRS 00:41:32.27 -49:36:19.60. In order to Nyquist sample the longest baselines we set pixel sizes of $0\rlap{.}{\arcsec}2$ and $0\rlap{.}{\arcsec}15$ for band 3 and 6 data, respectively. For each observation in ALMA band 3 and 6, we obtained two sets of ``dirty'' cubes (i.e., without performing the cleaning process) with a channel width of ${\rm 25\,km\,s^{-1}}$, and $40\,{\rm km\,s^{-1}}$, respectively in the LSB and USB. During the ``cleaning'' procedure, we set ``cube'' as spectral definition mode ({\tt specmode}) and ${\tt niter=0}$. We also obtained ``dirty'' continuum images by aggregating all the four SPWs in each frequency setting and we performed the Fourier transform with the {\tt tclean} by setting {\tt specmode=``mfs''}. The resulting beam sizes of the ALMA band 3 data are $1\rlap{.}{\arcsec}4\times 1\rlap{.}{\arcsec}3$, and $1\rlap{.}{\arcsec}5\times 1\rlap{.}{\arcsec}4$ for data cube at the reference frequency of $109.00\,{\rm GHz}$, and for continuum image, respectively. The data processing of ALMA band 6 observations yields to beam sizes of $1\rlap{.}{\arcsec}3\times 1\rlap{.}{\arcsec}1$, and $1\rlap{.}{\arcsec}2\times 1\rlap{.}{\arcsec}0$ for the continuum image, and the data cube at the reference frequency of $245.00\,{\rm GHz}$. After measuring the RMS (root-mean-square) of the ``dirty'' data and continuum images, we obtained ``cleaned'' data cubes and continuum images by setting tight circular apertures on $\ge 5\sigma$ sources in their continuum and we performed the cleaning using the {\tt tclean} task down to $2\sigma$ ({\tt nsigma=2}).

\subsection{Ancillary data}\label{ssect:ancillary_data}
As part of our ALMA program, the presented mosaics have been complemented with an additional single high-resolution ($\sim 0.25\arcsec$) pointing encompassing the central region of MQN01 field using ALMA band 3. We designed these observations in order to spatially resolve the CO(4--3) line emission of the quasar host galaxy CTS G18.01, as well as that of possible closely-separated galaxies. Full details and analysis of this data will be presented in a future paper. In the current work we benefit from this data to disentangle the QSO line and continuum emission which appear blended with a nearby companion located at $\sim 1\arcsec$ sky-projected distance in the lower-resolution mosaics (see, Sect.~\ref{sect:source_search}).

Our ALMA observations are part of an extensive multiwavelength observational campaign ranging from the (rest-frame) FIR to the X-ray regime using both ground-based and space telescopes. These data will be presented in full details in future works. In this work we benefit from some of the acquired data. Here we summarize the main characteristics of such observations. 

In this work, we make use of deep VLT/MUSE spectroscopic observations toward the MQN01 field (Cantalupo et al., in prep., and Galbiati et al., in prep.). Such observations consist in four 10 hours-pointing (40 hours of total exposure time) using the MUSE-WFM (Wide Field Mode) integral field spectrograph with adaptive optics covering a FoV of $\simeq  2\arcmin\times 2\arcmin$ which is fully sampled by our ALMA data. The MUSE observations provide us with integral field spectroscopy between $4650-9300\,{\rm \AA}$ with a spectral resolution of $R\sim 2000-3500$. In this field, galaxies are identified via their rest-frame far-UV (FUV) continuum emission in the white-light image. The final sample comprises only those sources with a high-confidence measurement of the spectroscopic redshift from interstellar absorption lines (e.g., ${\rm Si\RNum{2}}\,1260,1526\,{\rm \AA}$, ${\rm O\RNum{1}}\,1303\,{\rm \AA}$, ${\rm C\RNum{2}}\,1334\,{\rm \AA}$, and ${\rm C\RNum{4}}\,1548,1550\,{\rm \AA}$). The catalog includes 38 galaxies with secure redshift between $2.7<z<4.3$, out of which 22 lie within $\pm 1000\,{\rm km\,s^{-1}}$ with respect to the QSO CTS G18.01 systemic redshift ($z=3.25$; Galbiati et al., in prep., for full details). 

We imaged the MQN01 field in the optical by using the VLT/FORS2 \citep[Focal Reducer and low-dispersion Spectrograph 2;][]{Appenzeller+1992} instrument by acquiring broadband mosaics ($\simeq 13\times 13\,{\rm arcmin^2}$) in U (central wavelength $\lambda_0 = 361\,{\rm nm}$; $\sim 7\,{\rm hours}$ of exposure time), B ($\lambda_0 = 440\,{\rm nm}$; $\sim 5\,{\rm hours}$ of exposure time), and R ($\lambda_0 = 655\,{\rm nm}$; $\sim 45\,{\rm minutes}$ of exposure time) filters. Such observations cover a $36\times$ larger sky area with respect to the MUSE mosaic. This allows us to extend the census of the $z=3.0-3.5$ population of LBGs well beyond the MUSE FoV via color-color selection tested against the spectroscopic information available in the MUSE FoV (Galbiati et al. in prep.).

In this work, we additionally benefit from NIR photometric images taken with NIRCam (NIR Camera) instrument on board the {\it James Webb Space Telescope} \citep[JWST;][]{Rigby+2023}. Observations were acquired by using the extra-wide filters F150W2, and F322W2 in the short- and long-wavelength channel, respectively, with an on-source exposure time of $1632\,{\rm sec}$ per filter. These images cover a total FoV of $2\times5\,{\rm arcmin^{2}}$ across the two detectors of the camera, with one detector encompassing the sky area observed with MUSE. 

\begin{table*}[!t]
\def\arraystretch{1.15}
\caption{ALMA 1.2-mm continuum-selected sources.}  
\label{tbl:1.2mm_cand}      
\centering 
\resizebox{0.9\hsize}{!}{
\begin{tabular}{lcccccccc}     
\toprule\toprule
ID$_{\rm 1.2mm}$ (other ID) $\,^{(1)}$ & R.A.$\,^{(2)}$ & DEC.$\,^{(3)}$ & $S_{\rm 1.2\,mm}^{2\sigma}$$\,^{(4)}$ & $S_{\rm 1.2\,mm}^{\rm peak}$$\,^{(5)}$ & $S_{\rm 3\,mm}^{\rm peak}$$\,^{(6)}$ & ${\rm S/N}$$\,^{(7)}$ & fidelity$\,^{(8)}$ &$z_{\rm MUSE}$$\,^{(9)}$\\
	 	& (ICRS) & (ICRS) & (mJy) & $({\rm mJy\,beam^{-1}})$ & $({\rm \mu Jy\,beam^{-1}})$ & \\
\cmidrule(lr){1-9}
QSO CTS G18.01& 00:41:31.443 & -49:36:11.703 & $-$ & $0.99\pm0.04$ &  ${\rm 137 \pm 7}$ & $42.8$ & $1.00^{+0.00}_{-0.00}$ & $3.2365$ \\
Object B & 00:41:31.465 & -49:36:12.943 & $-$ & $1.63\pm0.04$ &  ${\rm 144 \pm 7}$ & $42.8$ & $1.00^{+0.00}_{-0.00}$ & $-$\\
C01 (L01)& 00:41:35.129 & -49:37:12.402 & ${\rm 2.0\pm0.2}$ & $0.75\pm0.04$ & ${\rm 77 \pm 7}$ & $20.0$ & $1.00^{+0.00}_{-0.00}$ & $3.2377$\\
C02 (L02)& 00:41:31.610 & -49:36:57.854 & ${\rm 0.9\pm0.2}$ & $0.69\pm0.04$ & ${\rm 57 \pm 7}$ & $19.7$ & $1.00^{+0.00}_{-0.00}$ & $3.2454$\\
C03 & 00:41:38.278 & -49:37:11.344 & ${\rm 0.9\pm0.4}$ & $0.59\pm0.06$ & {-} & $10.5$ & $1.00^{+0.00}_{-0.00}$ & $-$\\
C04 (L03)& 00:41:26.918 & -49:36:49.146 & ${\rm 0.5\pm0.3}$ & $0.42\pm0.04$ & ${\rm 30 \pm 7}$ & $9.9$ & $1.00^{+0.00}_{-0.00}$ & $-$\\
C05  & 00:41:35.513 & -49:35:24.551 & ${\rm 0.5\pm0.3}$ & $0.41\pm0.05$ & ${\rm 25 \pm 7}$ & $7.7$ & $1.00^{+0.00}_{-0.00}$ & $-$\\
C06 (i1) & 00:41:35.005 & -49:36:20.952 & $-$ & $0.21\pm0.04$ & ${\rm 39 \pm 7}$ & $5.7$ & $1.00^{+0.00}_{-0.00}$ & $2.5414^{(\dagger)}$\\
C07 & 00:41:36.858 & -49:37:09.698 & $-$ & $0.19\pm0.04$ & {-} & $4.7$ & $0.87^{+0.05}_{-0.05}$ & $-$\\
\cmidrule(lr){1-9}
C08 & 00:41:35.391 & -49:37:19.751 & $-$ & $0.22\pm0.05$ & {-} & $4.0$ & $0.5^{+0.2}_{-0.2}$ & $3.2539$\\
C09 (i5) & 00:41:37.999 & -49:36:43.895 & $-$ & $0.20\pm0.05$ & {-} & $4.1$ & $0.3^{+0.3}_{-0.3}$ & $2.8740$\\

\bottomrule
\end{tabular}
}
\tablefoot{{\it Upper part:} high-fidelity ($F\ge0.9$) candidate list extracted via blind search using {\sc LineSeeker}. {\it Lower part:} source candidates extracted by cross-matching ALMA 1.2-mm continuum-selected sources with $F> 0.2$ with the MUSE catalog of $z\ge2.5$ sources. $^{(1)}$Identifier of ALMA 1.2-mm continuum-selected candidates. In case a source is also detected in its CO(4--3) line (see, Sect.~\ref{ssect:co43-extraction}), the corresponding identifier from Table~\ref{tbl:CO_cand} is reported. In case a source has been identified as low-$z$ interloper, its identifier is also reported (see, Sect~\ref{ssect:interlopers}). The QSO and the nearby companion are labeled with the Cal\'an-Tololo Survey \citep[CTS,][]{Maza+1995} identifier, and with ``Object B'', respectively. Object B has been selected by {\sc LineSeeker} together with the QSO as a single source. Here we report, separately the coordinate of the continuum peak of the QSO CTS G18.01 and Object B obtained via a visual inspection. Their ${\rm S/N}$ and the fidelity are set to the values provided by the code that are referred to the continuum peak of the Object B which appears brighter than the QSO at $1.2\,{\rm mm}$. $^{(2)}$Right Acension (ICRS). $^{(3)}$Declination (ICRS). $^{(4)}$Integrated flux density over the $\ge2\sigma$ isophote. {\rm This quantity is not reported for compact sources at low S/N and for the QSO and Object B which are partially blended.}  $^{(5)}${\rm Continuum peak flux density at $1.2\,{\rm mm}$}.  {\rm $^{(6)}$ Continuum peak flux density at $3\,{\rm mm}$}. $^{(7)}${\rm Signal-to-noise ratio of the 1.2mm-selected sources}. $^{(8)}$Fidelity and its uncertainties estimated by {\sc LineSeeker} using negative detections. $^{(9)}$High-confidence redshift estimate from MUSE spectroscopic data. For those sources for which the redshift is not provided they either have an uncertain redshift measurement or they do not have any FUV-continuum counterpart from MUSE. $^{(\dagger)}$Source C06 has been unambiguously identified as a low-$z$ interloper (``i1'') showing a bright emission line in the LSB of 3-mm ALMA band. For this galaxy we therefore report the low-confidence redshift provided by the analysis of the MUSE spectrum.}
\end{table*}

\section{Source search and characterization}\label{sect:source_search}
We performed a blind search of continuum- and line-emitting sources in both our ALMA band 3 and 6 images and cubes by using the Python-based code {\sc LineSeeker}\footnote{The code is publicly available at the following link: \url{https://github.com/jigonzal/LineSeeker}} \citep[see, ][for full details]{Gonzalez-Lopez+2017, Gonzalez-Lopez+2019}. This code was originally developed to search for line and continuum emission of galaxies in the ALMA Frontier Fields survey \citep[see][]{Gonzalez-Lopez+2017b, Gonzalez-Lopez+2017}, and was subsequently employed in various other surveys such as the ALMA Spectroscopic Survey in the Hubble Ultra Deep Field \citep[ASPECS; see, e.g.,][]{Decarli+2019c, Gonzalez-Lopez+2019, Gonzalez-Lopez+2020}, the Multiwavelength Study of ELAN Environments \citep[AMUSE$^{2}$; see, e.g.][]{Chen+2021, Arrigoni-Battaia+2022}, and the Northern Extended Millimeter Array (NOEMA) Molecular Line Scan of the Hubble Deep Field North \citep{Boogaard+2023}. Here we summarize the operation and the main features of the code.

{\sc LineSeeker} adopts a matched-filter approach. The code combines adjacent spectral channels by convolving the data cube along the spectral axis using Gaussian kernels with a range of widths. The RMS of the resulting images is then estimated via a sigma clipping at $5\sigma$ to remove the spurious increases of the noise due to possibly bright lines or continuum emission within the convolved channels. Then all the voxels above a given signal-to-noise ratio\footnote{{\sc LineSeeker} estimates the S/N of the source candidates on the basis of the peak flux density per beam.} (${\rm S/N}$) threshold are stored for each convolution kernels. Finally, a list of line (or continuum\footnote{Thanks to its design, {\sc LineSeeker} can also be used to perform 2D source search in images by simply skipping channel convolution steps.}) emitter candidates is generated by grouping voxels from the different channels corresponding to unique sources by using the Density-Based Spatial Clustering of Application with Noise algorithm \citep[{\sc DBSCAN};][]{Ester+1996} included in the Python package Scikit-learn \citep{Pedregosa+2011}. The final ${\rm S/N}$ of the sources is then selected as the maximum value obtained through the different convolutions. 
{\sc DBSCAN} is also able to recover extended sources traced by ${\rm S/N}\ge2$ pixels, however, a visual inspection is needed in order to verify if the single extended source is actually composed by multiple blended sources along the line of sight.

For each source candidate selected by {\sc LineSeeker}, the code automatically estimates the probability of false-positive detection based on the source ${\rm S/N}$. To this purpose, the code is run on the negative (i.e., multiplied by $-1$) cube or image. In a deep extragalactic field in the (sub-)mm, the majority of the surveyed area is expected to be empty sky; hence, any ``negative'' peak is a realization of noise. The statistics of negative detections is then compared to that of the positive ones. The fidelity (or reliability) of a positive detection as a function of its ${\rm S/N}$ is therefore computed as

\eq{F\,({\rm S/N}) = 1 - \frac{N_{\rm neg}({\rm S/N})}{N_{\rm pos}({\rm S/N})},\label{eq:fidelity}}

where $N_{\rm neg}$ and $N_{\rm pos}$ are the number of negative and positive detections at a given ${\rm S/N}$, respectively. To compute the fidelity at any ${\rm S/N}$ following Eq.~(\ref{eq:fidelity}), {\sc LineSeeker} assumes that the noise in the data is Gaussian distributed and compute the best fit model of the cumulative distribution of negative detections using a function of the form $N[1-{\rm erf}({\rm SN}/\sqrt{2}\,\sigma)]$, with {\rm erf} that is the error function and $N$, $\sigma$ are free parameters.

Other similar source-finding algorithms are available in the literature such as {\sc FindClump} \citep{Walter+2016}, and {\sc MF3D} \citep{Pavesi+2018}, that mostly differ on details (such as, e.g., the adopted spectral filter function). Comparisons between the codes yield consistent results to within $\sim 10\%$ \citep[see,][]{Gonzalez-Lopez+2019}.

\subsection{1.2-mm continuum-selected candidates}\label{ssect:1.2mm-extraction}
We performed a source search of 1.2-mm continuum-emitting galaxy candidates in MQN01 field by running {\sc LineSeeker} on the ``dirty'' continuum band 6 image, excluding the region with PB response $<50\%$ in which the low telescope sensitivity enhances the fraction of spurious candidates. The ``dirty'' data are preferred over the ``cleaned'' ones since in the former the intrinsic properties of the noise are preserved. Also, we do not correct our ``dirty'' image for the PB response to preserve the spatial homogeneity of the noise across the FoV. We therefore extracted all ${\rm S/N}\ge 3$ detections, and we selected the source candidates with estimated fidelity of $F\ge 90\%$ corresponding to ${\rm S/N}\ge 4.7$. With this method, we retrieve a total of nine sources including the QSO CTS G18.01, and a closely (on-sky) separated source (hereafter, Object B) partially blended with the QSO.

\begin{figure*}[!htbp]
   	\centering
   	\resizebox{\hsize}{!}{
		\includegraphics{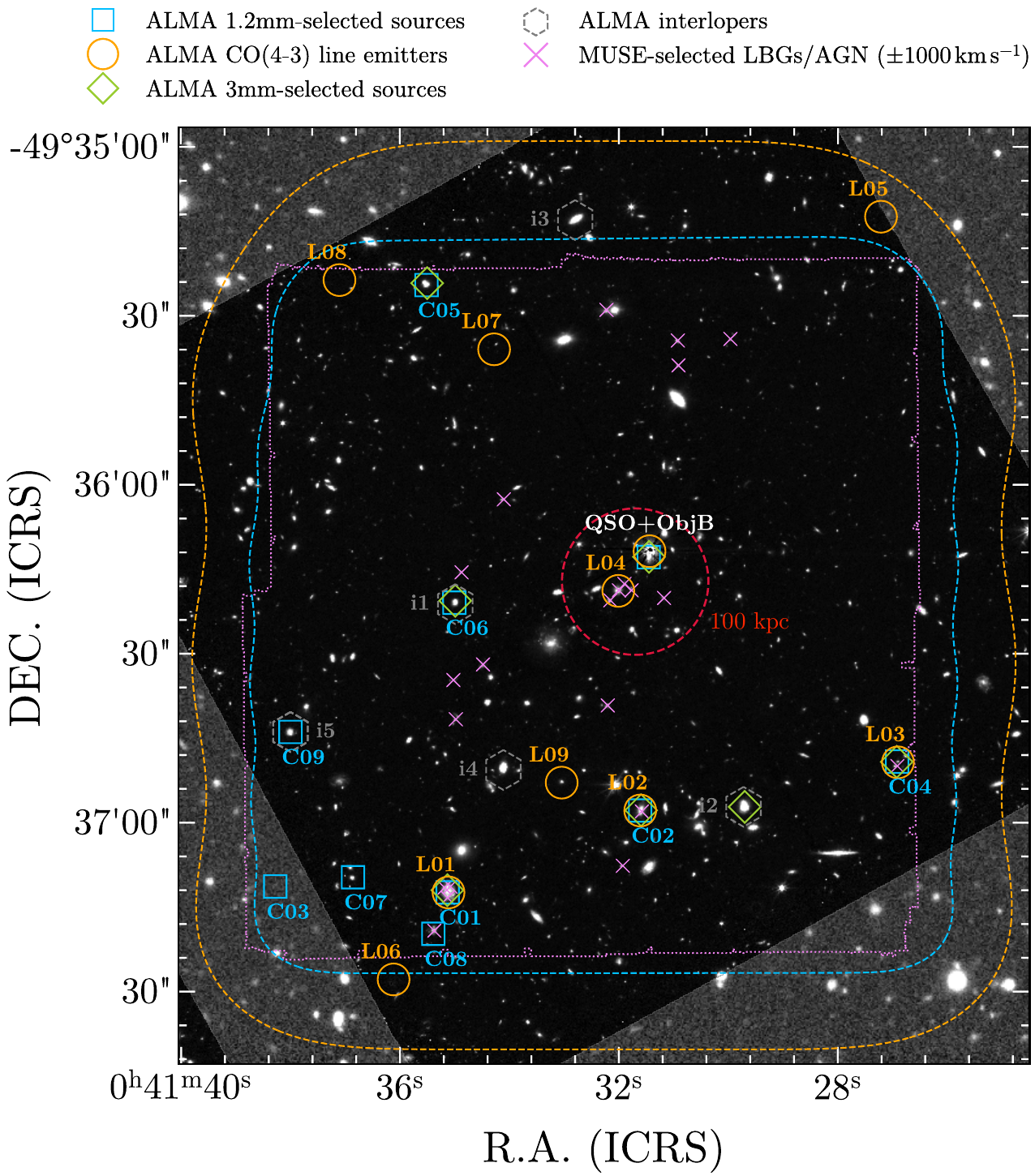}}
       \caption{Footprints of our ALMA and MUSE observations toward the MQN01 field. The orange and blue contours encircle the area where the combined mosaic PB response is $\ge 0.5$ for ALMA band 3 and 6, respectively. Within these areas we performed the source search. Violet contour draws the MUSE footprint. The background is a composition of the JWST NIRCam F322W2 image (center and south-est corner) complemented with the VLT/FORS2 $R$-band observation (north-est, north- and south-west corners, and the south-est gap between the JWST NIRCam detectors). The PSF-like emission of QSO CTS G18.01 has been subtracted in the JWST image revealing a nearby south-est quasar companion (Object B). Orange circles and blue squares indicate the locations of the ALMA CO(4--3) line-emitting and 1.2-mm continuum-selected sources in our ALMA band 3 and 6 observations, respectively. Green squares pin-point galaxy candidates detected in continuum at $3\,{\rm mm}$ in ALMA band 3. Grey hexagons denote sources corresponding to low-$z$ counterparts, all but i5 show bright emission line in ALMA band 3. Violet crosses are the MUSE-selected LBGs belonging within $\pm 1000\,{\rm km\,s^{-1}}$ with respect to the QSO CTS G18.01 systemic redshift. The dashed red circle shows the inner $100\,\rm pkpc$ around the estimated center of the protocluster core.}
       \label{fig:selections}
   \end{figure*}

We complemented our blind search of 1.2-mm continuum candidates in the field by cross-matching our MUSE catalog of high-$z$  sources ($z>2.5$, see Sect.~\ref{ssect:ancillary_data}) with the low-fidelity ($F> 20\%$) sources selected by {\sc LineSeeker}. In this process, we cross-matched the on-sky spatial position of the MUSE and ALMA continuum sources within a separation limit of $0\rlap{.}{\arcsec}6$. We chose this separation since it is about one half of the angular resolution of the ALMA image thus accounting for possible spatial offset between the ALMA low-${\rm S/N}$ FIR- and MUSE FUV-continuum peak\footnote{Such spatial offsets can be produced due to noise fluctuations or differential dust obscuration.}. This separation also corresponds to the maximum observed angular distance between our ALMA candidates selected in the blind search and their MUSE counterparts. By doing so, we recovered two additional sources in the field. In Table~\ref{tbl:1.2mm_cand} we report the final catalog of the eleven ALMA 1.2-mm continuum-selected sources. In Fig.~\ref{fig:selections}, we show the location of the sources detected in MUSE within $\pm 1000\,{\rm km\,s^{-1}}$ with respect to QSO CTS G18.01 and the ALMA 1.2-mm continuum-selected sources in the MQN01 field. We labeled the latter as C01 -- C09.

\subsection{CO(4--3) line-emitting candidates}\label{ssect:co43-extraction}
We used {\sc LineSeeker} to blindly search for CO(4--3) emission lines that are expected to be redshifted in the USB of the ALMA band 3 datacube. We opted to extract sources in the ``dirty'' cube not corrected for the PB response over the area where the combined mosaic sensitivity is $\ge 50\%$ (see also, Sect.~\ref{ssect:1.2mm-extraction}). We run the line-search algorithm on the datacube that we spectrally binned at $25\,{\rm km\,s^{-1}}$ using Gaussian kernels with widths ranging from 0 to 18 channels. This range enables the code to match the typical linewidths of CO lines observed in high-$z$ galaxies (${\rm FWHM} \sim 50-1000\,{\rm km\,s^{-1}}$; see, e.g., \citealt{CarilliWalter2013}). We extracted all line-emitting candidates with estimated ${\rm S/N}\ge 3$. We then selected those sources with estimated fidelity of $F\ge 90\%$ (see, Sect.~\ref{sect:source_search}). Similarly to the search of 1.2-mm continuum emitters, we cross-matched the location of the MUSE LBGs with the low-fidelity ($F>20\%$) line-emitting candidates from {\sc LineSeeker} within a separation limit of $0\rlap{.}{\arcsec}6$. As a result of this procedure, we extracted one additional source. The CO-based redshift of this candidate is within $\pm 200\,{\rm km\,s^{-1}}$ with that based on the Ly-$\alpha$ emission line derived from MUSE spectroscopic data. The difference is consistent with the typical shift observed between the Ly-$\alpha$ line and the systemic redshift of high-$z$ Ly-$\alpha$ emitters \citep[LAEs; see, e.g.,][]{Guaita+2013, Muzahid+2020, Muzahid+2020b, Muzahid+2021}. Within the selected sample of 13 galaxies we identified two interlopers at lower redshift (dubbed as ``i3'' and ``i4'') which we therefore excluded from the final sample (see, Sect.~\ref{ssect:interlopers}, for a detailed discussion). In Table~\ref{tbl:CO_cand} we report the final catalog of the eleven ALMA CO(4--3) line-emitting candidates. In Fig.~\ref{fig:selections}, we draw the location of these galaxies (L01 -- L09) as well as that of the identified low-$z$ interlopers within the field.

\begin{table*}[!htbp]
\def\arraystretch{1.15}
\caption{ALMA CO(4--3) line-selected sources.}  
\label{tbl:CO_cand}      
\centering 
\resizebox{\hsize}{!}{
\begin{tabular}{lccccccccccc}     
\toprule\toprule
ID$_{\rm CO}$ (ID$_{\rm 1.2mm}$)$\,^{(1)}$ & R.A.$\,^{(2)}$ & DEC.$\,^{(3)}$ & $F_{\rm CO}$$\,^{(4)}$ & FWHM$_{\rm CO}$$\,^{(5)}$  & $L_{\rm CO}$$\,^{(6)}$ & $L'_{\rm CO}$$\,^{(7)}$ & ${\rm S/N}$$\,^{(8)}$ & fidelity$\,^{(9)}$ &  $z_{\rm CO}$$\,^{(10)}$ &$\Delta\varv_{\rm QSO}$$\,^{(11)}$ &$z_{\rm MUSE}$$\,^{(12)}$ \\
						 & (ICRS) & (ICRS) & $({\rm Jy\,km\,s^{-1}})$ & $({\rm km\,s^{-1}})$ & $(10^{8}\,L_{\odot})$ & $(10^{10}\,{\rm K\,km\,s^{-1}\,pc^{2}})$ & & & & $({\rm km\,s^{-1}})$\\
\cmidrule(lr){1-12}
QSO CTS G18.01		& 00:41:31.443 & -49:36:11.703 & $0.86^{+0.08}_{-0.07}$ & $560^{+56}_{-51}$ & $0.80^{+0.07}_{-0.07}$ & $2.6^{+0.2}_{-0.2}$ & $35.9$ & $1.00^{+0.00}_{-0.00}$ &$3.2502^{+0.0003}_{-0.0003}$ &$0\pm20$ &$3.2365$\\
Object B 				& 00:41:31.463 & -49:36:12.943 & $1.25^{+0.10}_{-0.10}$ & $753^{+62}_{-55}$ & $1.16^{+0.10}_{-0.09}$ & $3.7^{+0.3}_{-0.3}$ & $35.9$ & $1.00^{+0.00}_{-0.00}$ &$3.2475^{+0.0004}_{-0.0004}$ &$-191\pm26$ &$-$\\
L01 (C01)& 00:41:35.113 & -49:37:12.402 & $1.41^{+0.09}_{-0.09}$ & $424^{+31}_{-28}$ & $1.31^{+0.09}_{-0.08}$ & $4.2^{+0.3}_{-0.3}$ & $25.1$ & $1.00^{+0.00}_{-0.00}$ 				 &$3.24509^{+0.00019}_{-0.00018}$ &$-362\pm12$ &$3.2377$\\
L02 (C02)& 00:41:31.610 & -49:36:57.854 & $0.79^{+0.11}_{-0.11}$ & $939^{+134}_{-123}$ & $0.73^{+0.10}_{-0.10}$ & $2.3^{+0.3}_{-0.3}$ & $13.7$ & $1.00^{+0.00}_{-0.00}$ 				 &$3.25081^{+0.0009}_{-0.0009}$ &$42\pm60$ &$3.2454$\\
L03 (C04) & 00:41:26.902 & -49:36:49.296 & $0.34^{+0.05}_{-0.04}$ & $548^{+79}_{-70}$ & $0.32^{+0.04}_{-0.04}$ & $1.01^{+0.14}_{-0.13}$ & $10.0$ & $1.00^{+0.00}_{-0.00}$ & $3.2494^{+0.0005}_{-0.0005}$ &$-54\pm36$ & $-$\\
L04 & 00:41:32.011 & -49:36:18.854 & $0.13^{+0.03}_{-0.02}$ & $239^{+55}_{-45}$ & $0.12^{+0.02}_{-0.02}$ & $0.38^{+0.07}_{-0.07}$ & $6.5$ & $1.00^{+0.00}_{-0.00}$ & $3.2456^{+0.0003}_{-0.0003}$ &$-328\pm25$ &$3.2430$\\
L05 & 00:41:27.214 & -49:35:12.397 & $0.29^{+0.09}_{-0.07}$ & $667^{+353}_{-198}$ & $0.27^{+0.08}_{-0.07}$ & $0.9^{+0.3}_{-0.2}$ & $5.8$ & $0.97^{+0.01}_{-0.01}$ 					 &$3.2999^{+0.0011}_{-0.0014}$ &$3460\pm83$ &$-$\\
L06 & 00:41:36.117 & -49:37:27.85 & $0.19^{+0.04}_{-0.04}$ & $390^{+101}_{-84}$ & $0.18^{+0.04}_{-0.04}$ & $0.56^{+0.13}_{-0.12}$ & $5.6$ & $0.90^{+0.02}_{-0.02}$ 				 &$3.2736^{+0.0006}_{-0.0006}$ &$1640\pm43$ &$-$\\
L07 & 00:41:34.279 & -49:35:35.953 & $0.11^{+0.03}_{-0.03}$ & $253^{+60}_{-59}$ & $0.10^{+0.03}_{-0.03}$ & $0.31^{+0.08}_{-0.08}$ & $5.6$ & $0.88^{+0.04}_{-0.04}$ 				 &$3.2231^{+0.0005}_{-0.0005}$ &$-1928\pm36$ &$-$\\
L08 & 00:41:37.102 & -49:35:23.648 & $0.24^{+0.07}_{-0.06}$ & $701^{+237}_{-231}$ & $0.22^{+0.06}_{-0.06}$ & $0.71^{+0.20}_{-0.18}$ & $5.5$ & $0.88^{+0.04}_{-0.04}$ 				 &$3.2281^{+0.0011}_{-0.0010}$ &$-1571\pm77$ &$-$\\
\cmidrule(lr){1-12}
L09 & 00:41:33.045 & -49:36:52.904 & $0.11^{+0.03}_{-0.03}$ & $325^{+118}_{-78}$ & $0.10^{+0.03}_{-0.03}$ & $0.33^{+0.09}_{-0.08}$ & $4.7$ & $0.3^{+0.3}_{-0.3}$ 					 &$3.2440^{+0.0006}_{-0.0006}$ &$-440\pm41$ &$3.2452$\\
\bottomrule
\end{tabular}
}
\tablefoot{{\it Upper part:} high-fidelity ($F\ge0.9$) candidate list extracted via blind search using {\sc LineSeeker}. {\it Lower part:} source candidates extracted by cross-matching ALMA CO(4--3) line-emitting candidates with $F> 0.2$ with the MUSE catalog of $z\ge2.5$ sources. $^{(1)}$Identifier of ALMA CO(4--3) line-emitting (1.2-mm continuum-selected) candidates. The QSO and the nearby companion (Object B) are labeled as in Table~\ref{tbl:1.2mm_cand}. Object B has been selected by {\sc LineSeeker} together with the QSO as a single source. Here we report, separately the coordinate of the continuum peak of the QSO CTS G18.01 and Object B obtained via a visual inspection. Their ${\rm S/N}$ and the fidelity are set to the values provided by the code that are referred to the CO(4--3) peak of the QSO. $^{(2)}$Right Acension (ICRS). $^{(3)}$Declination (ICRS). $^{(4)}$Our best estimate of the CO(4--3) flux derived from the Gaussian fit of the line emission (see Sect~\ref{ssect:flux_measure}). $^{(5)}$Full-width-at-half-maximum of the CO line. $^{(6),\,(7)}$CO(4--3) luminosities estimated via Eq.~(\ref{eq:lum}) and~(\ref{eq:lum1}), respectively. $^{(8)}$Maximum value of the Signal-to-noise ratio obtained through the different convolution (see Sect.~\ref{sect:source_search}). $^{(9)}$Fidelity and its uncertainties estimated by {\sc LineSeeker} using negative detections. $^{(10)}$Redshift estimate from the CO line centroid. $^{(11)}$Velocity shift with respect to the QSO CTS G18.01 redshift. $^{(12)}$High-confidence redshift estimate from MUSE spectroscopic data. For those sources for which the redshift is not provided they either have an uncertain redshift measurement or they do not have any FUV-continuum counterpart from MUSE.}
\end{table*}

\subsection{3-mm continuum-selected candidates}\label{ssect:3mm-inter-extraction}
Similarly to what described in Sect.~\ref{ssect:1.2mm-extraction}, we complemented our source extraction by searching candidates detected in continuum at $3\,{\rm mm}$. For this purpose, we run {\sc LineSeeker} on the ``dirty'' aggregate (LSB+USB) continuum ALMA band 3 image. This blind search results in six continuum-detected sources (including the QSO and Object B) with fidelity $F\ge 90\%$ . We also carefully inspected the ALMA $3\rm{mm}$-continuum image in the position of secure sources ($F=100\%$) revealed in the ALMA band 6 at $1.2\,{\rm mm}$. We therefore included two additional sources in the final sample corresponding to C04 and C05 (see, Table~\ref{tbl:1.2mm_cand}) which show convincing continuum emission at $3\,{\rm mm}$. We verified our conclusion by cross-matching the catalog of secure positive continuum detections at $1.2\,{\rm mm}$ with that at $3\,{\rm mm}$ provided by {\sc LineSeeker}. Finally, the cross-match between the catalog of high-$z$ MUSE LBGs and $3\,{\rm mm}$-continuum candidates did not provide us with any additional source. The locations of these detections in the MQN01 field are indicated in Fig.~\ref{fig:selections} by green squares. 

\subsection{Completeness and Flux Boosting}\label{ssect:complteness}
In order to compute the source number counts and the CO(4--3) LF it is crucial to determine the probability of detecting sources in our blind search. Such information is enclosed in the continuum and CO line completeness functions. In addition, we need to estimate how the measured fluxes are affected by the noise in the real data. Indeed, sources with low ${\rm S/N}$ have higher probability to be recovered with higher flux than the intrinsic value because of the boosting effect produced by the noise fluctuations \citep[the so-called ``flux-boosting effect''; see, e.g.,][]{Hogg+1998, Scott+2002, Coppin+2006}. We computed the completeness of our survey by following a common approach widely adopted in the literature \citep[see, e.g.,][]{Hatsukade+2016, Hatsukade+2018, Umehata+2017, Gonzalez-Lopez+2019, Gonzalez-Lopez+2020, Bethermin+2020, Boogaard+2023}. We injected artificial sources in the real data and we then performed the source search by using \textsc{LineSeeker}. More details about this exercise and how it is used to perform the corrections to the LFs are described in the following.

\begin{figure}[!t]
   	\centering
   	\resizebox{\hsize}{!}{
      		\includegraphics{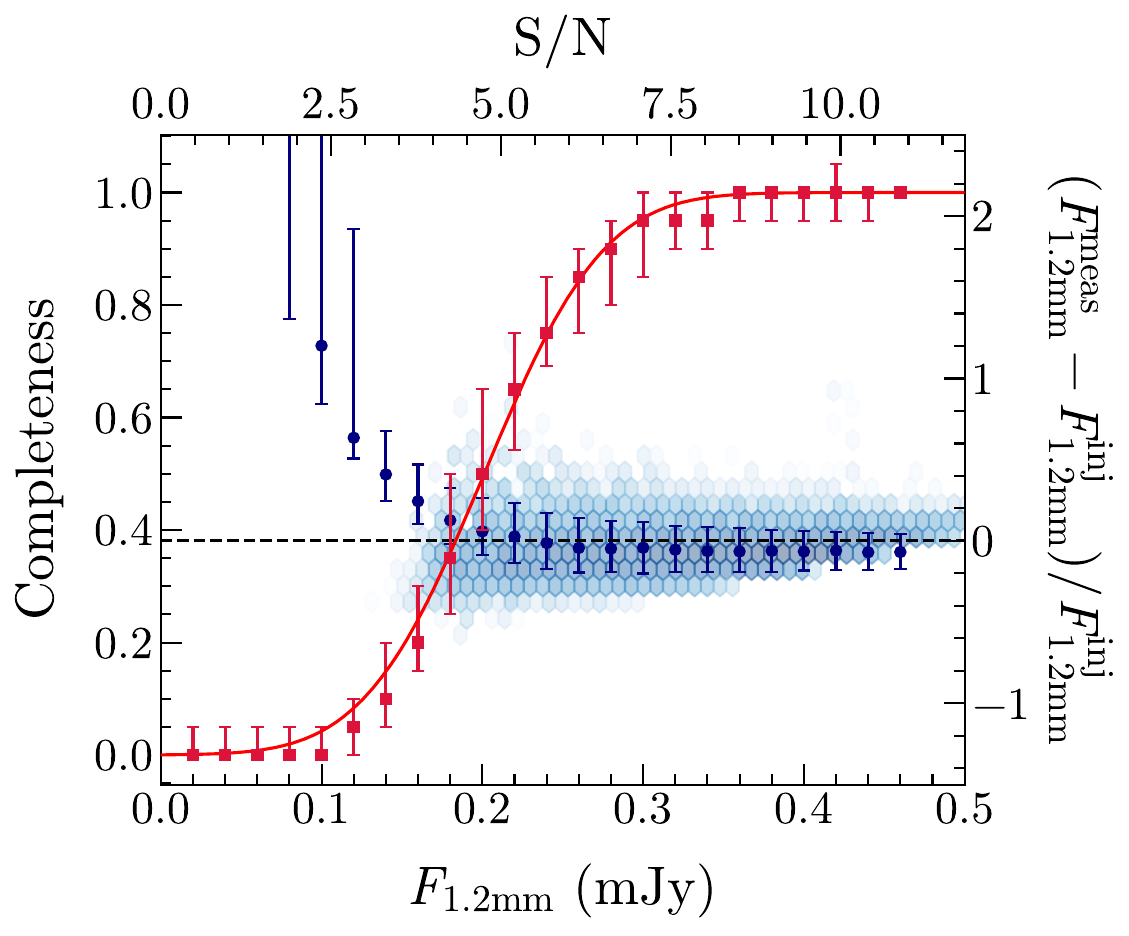}}
       \caption{Completeness (red squares, {\it left axis}) and flux boosting (blue circles and density plot, {\it right axis}) corrected for the PB response as a function of 1.2-mm flux density ({\it bottom axis}) and ${\rm S/N}$ ({\it top axis}) of injected sources in the ALMA band 6 continuum image. The errorbars are derived by computing the 16th and 84th percentile of the distributions of completeness and flux boosting measurements in each flux bin. The red solid line is the best-fitting function to the completeness values modeled as $C(F_{\rm 1.2\,mm})=\{1+{\rm erf}[(F_{\rm 1.2\,mm} - A)/B]\}/2$ with $A=0.197\pm0.004$ and $B=0.083\pm0.007$. The density plot shows the values of the flux boosting as a function of the measured flux of the injected sources.}
       \label{fig:12mm_completeness}
\end{figure}

To estimate the completeness function of $1.2\,{\rm mm}$-continuum detections in the ALMA band 6 we created artificial point-like sources by rescaling the synthesized beam model and we injected them into the dirty ALMA Band 6 continuum map at random positions within the source-search area of our survey (i.e., where the combined mosaic PB response is $\ge 50\%$). To take into account the effect of the variation of sensitivity across the mosaic field, we rescaled the source fluxes by the PB response at the input location of each source. We then run the source-search algorithm and we computed the source detection rate. We considered a source recovered if it is extracted within $1\arcsec$ from its input location with a fidelity $\ge 90\%$. We repeated this procedure by injecting $20$ sources simultaneously in the image and iterating for $100$ times for each 1.2-mm flux density value within the range $0.02-0.46\,{\rm mJy}$ in steps of $0.02\,{\rm mJy}$. The total number of injected sources in the simulation is $46000$. During the simulation we also evaluated the effect of flux boosting by computing the ratio of measured-to-injected flux of the artificial sources as $(F^{\rm meas}_{\rm 1.2\,mm}-F^{\rm inj}_{\rm 1.2\,mm})/F^{\rm inj}_{\rm 1.2\,mm}$. In Fig.~\ref{fig:12mm_completeness} we report the output of our simulation. As a result, the completeness in the flux range of the detected sources is in the range $50-100\%$, while the flux boosting effect does not affect significantly the measured 1.2-mm flux density of our selected sample of galaxies.

\begin{figure}[!t]
   	\centering
   	\resizebox{\hsize}{!}{
      		\includegraphics{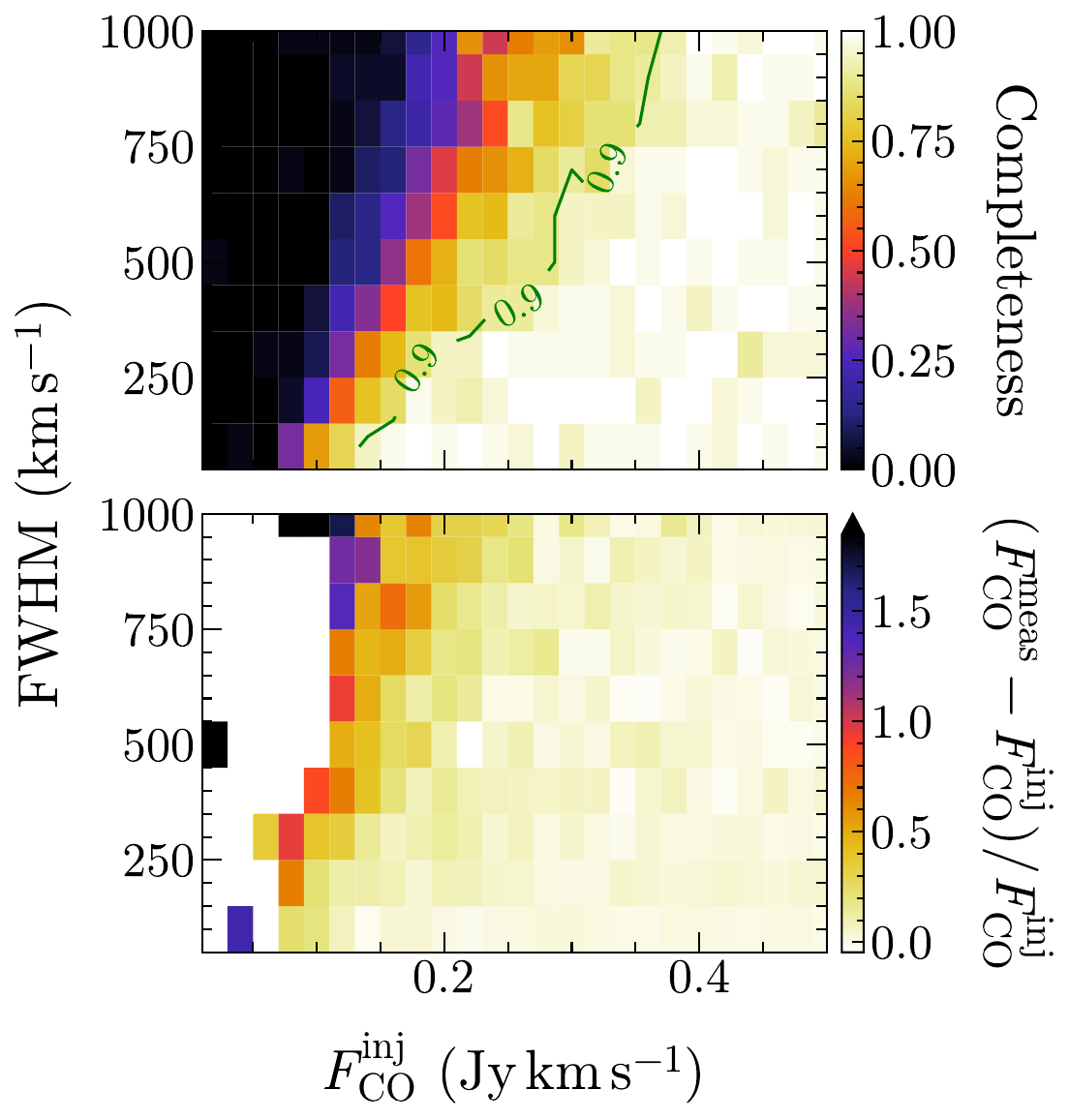}}
       \caption{Completeness ({\it top panel}) and flux boosting ({\it bottom panel}) corrected for the PB response as a function of the line-velocity integrated flux and the line FWHM of the injected sources in the ALMA band 3 dirty cube binned at $25\,{\rm km\,s^{-1}}$. In the top panel, the green line indicate where the completeness is $\ge 90\%$.}
       \label{fig:compl_fb}
\end{figure}

We evaluated the completeness of our blind survey of line-emitting sources and the effect of the flux boosting on the observed line emission. On the basis of the line measurements of the selected CO(4--3) emitters, we injected point-like artificial emission lines using Gaussian spectral profile with FWHM and velocity-integrated flux ranging within $100-1000\,{\rm km\,s^{-1}}$ in steps of $100\,{\rm km\,s^{-1}}$, and $0.02-0.5\,{\rm Jy\,km\,s^{-1}}$ in steps of $0.02\,{\rm Jy\,km\,s^{-1}}$, respectively. We injected the artificial sources at random positions (within the volume where the mosaic PB response is $\ge 50\%$) in the USB of the dirty ALMA band 3 cube binned at ${\rm 25\,km\,s^{-1}}$ (i.e., where we search for the CO(4--3)-line emitting candidates; see~Sect.\ref{ssect:co43-extraction}). We then rescaled the signal from the artificial line emitters in each channel of the cube by the PB response at the input position of the sources. We then ran \textsc{LineSeeker} and we computed the source detection rate following the same criteria adopted for the artificial sources in the continuum. To estimate the line flux boosting effect for the recovered sources, we obtained the line-velocity integrated map using channels within $\pm2\sigma$ with respect to the line centroid provided by \textsc{LineSeeker}\footnote{Here we assumed that the recovered line is a Gaussian with the input FWHM.}. We then measured the total source fluxes by performing fit of the moment-0 map by using a 2D Gaussian model\footnote{This method is equivalent of performing Gaussian fit of the line profile along the spectral axis but has the advantage of being more robust against fit failures. It is therefore preferable to estimate the fluxes of a large number of sources.}. For each values of the line FWHM and line-velocity integrated flux, we injected simultaneously 50 sources in the cube for a total of 12500 sources injected in the whole simulation. In Fig.~\ref{fig:compl_fb} we report the output of our simulation. As a result the selected sample is complete for line fluxes $\gtrsim 0.4\,{\rm Jy\,km\,s^{-1}}$ while the flux boosting effect has negligible impact on the line measurements of our selected sample of CO(4-3) line-emitting galaxies.

\subsection{Identification of interlopers across the redshift range}\label{ssect:interlopers}
\begin{figure}[!t]
   	\centering
   	\resizebox{\hsize}{!}{
      		\includegraphics{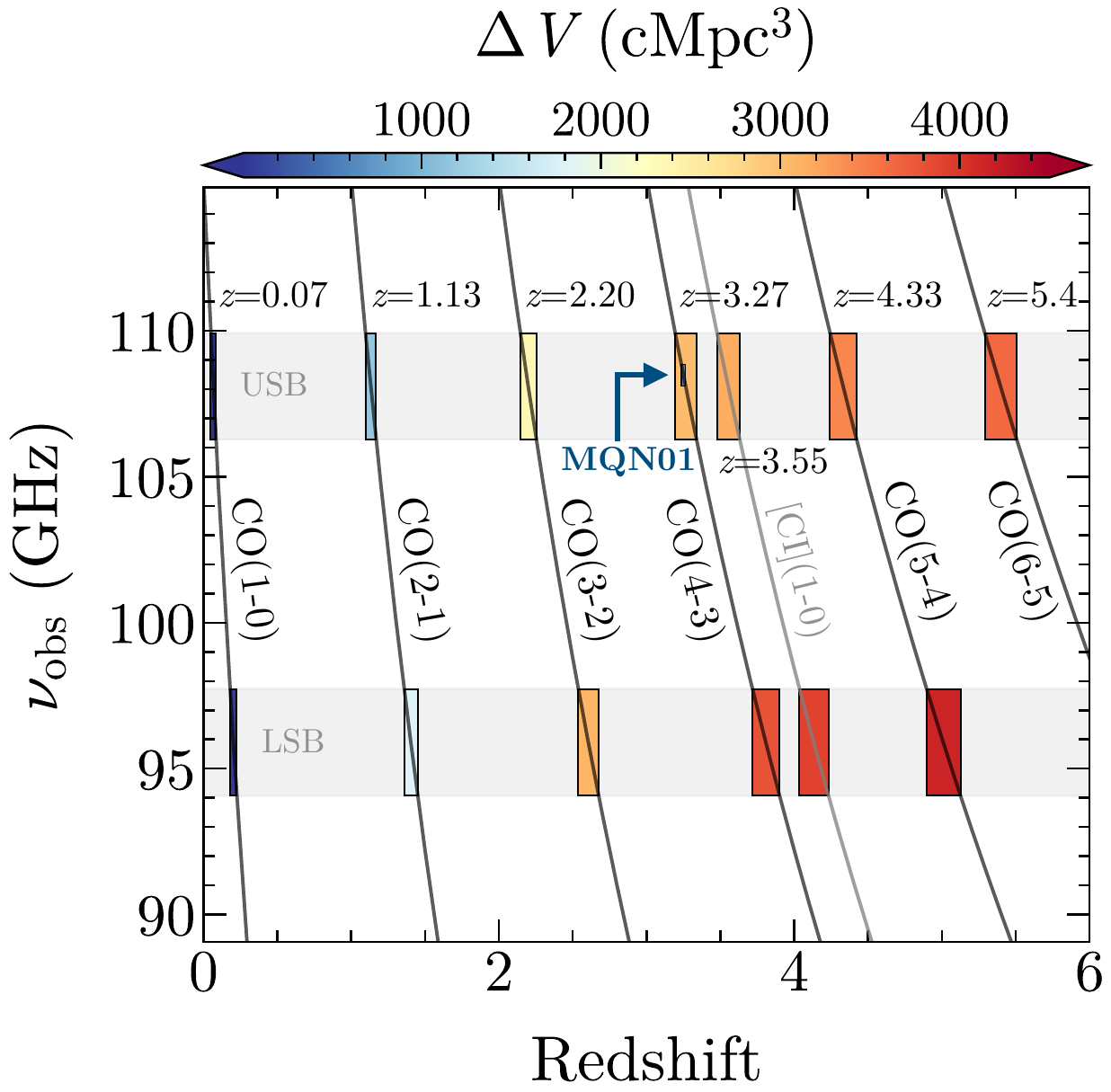}}
       \caption{The observed frequency of CO and \ci{609} transitions as a function of redshift. The boxes encircle both the frequency and redshift ranges covered by our ALMA band 3 observations. Boxes are color-coded by the cosmological volume probed by each transitions. Grey bands show the SPW coverage in the lower (LSB) and the upper sideband (USB). The mean redshift of transitions entering in the USB are also reported. The blue arrow points to the box representing the volume within $\pm 1000\,{\rm km\,s^{-1}}$ around the QSO CTS G18.01 systemic redshift.}
       \label{fig:interlopers}
   \end{figure}

For the purpose of our analysis it is crucial to identify possible interlopers at different redshifts among the ALMA-selected galaxies. In the ALMA 3-mm band, sources at any redshift are expected to be revealed in their line emission primarily via CO rotational lines, or the fine-structure transitions of the atomic neutral carbon \ci{609,\,370}\ \citep[albeit the latter are expected to be much fainter than low-$J$ CO lines; see, e.g.,][but see, also, \citealt{Gullberg+2016}]{Gonzalez-Lopez+2019, Decarli+2020}. In Fig.~\ref{fig:interlopers}, we draw the redshifted frequency of such transitions entering in our ALMA band 3 SPWs up to $z=6$. The various transitions probe galaxies within different redshift intervals and cosmological volumes depending on the surveyed sky area and the encompassed frequency range (see, Table~\ref{tbl:interlopers}). By using the available best-fit model to the CO LFs from \citet{Boogaard+2023}, we computed the number of expected galaxies at various redshifts within the cosmological volume probed by our observations above the limiting luminosity (see, Table~\ref{tbl:interlopers}). As a result, $\sim 2$ sources at $z\sim1.1$ are expected to be detected via their CO(2--1) line in the USB of our ALMA band 3 survey, $\sim 1$ at $z\sim2.2$ through CO(3--2) line, and $\sim 0.5$ CO(5--4) line emitters at $z\sim4.3$. For the targeted CO(4--3) line the number of expected sources is $\sim2.5$, dropping to $\sim 0.5$ when considering the volume within $\pm 1000\,{\rm km\,s^{-1}}$ around the QSO CTS G18.01. However, these estimates suffer from large uncertainties given the poor sampling of the CO LFs \citep[see, e.g.,][]{Decarli+2019c,Decarli+2020, Boogaard+2023}, and therefore must be taken with caution. 

In order to unambiguously identify the emission lines observed in our survey, we made use of our multiwavelength datasets of the MQN01 field. As described in Sect.~\ref{ssect:co43-extraction}, our blind search for CO(4--3) line-emitting candidates yielded an initial sample of 13 sources. Within the candidates belonging to this sample, 4 of out 13 ($\simeq30\%$; L03, L07, L08, i4, excluding Object B) do not correspond to any high-$z$ MUSE-selected LBGs/AGN. In addition, three candidates (L05, L06, i3) are located outside the MUSE mosaic footprint. For the first group of galaxies, the analysis of their MUSE spectra enabled us to identify i4 as a low-$z$ interloper. The spectrum of this galaxy shows ${\rm Mg\RNum{2}]\,2796, 2803\,\AA}$ absorptions, and  ${\rm [O\RNum{2}]\,3726,3729\,\AA}$ line emission, the wavelengths of which unambiguously place this object at $z\simeq1.170$, hence the observed line emission in our ALMA band 3 observation is the CO(2--1) at the redshifted frequency of $106.2\,{\rm GHz}$. In the case of sources outside the area surveyed by MUSE, the analysis of UBR colors obtained from VLT/FORS2 observations (see, Sect.~\ref{ssect:ancillary_data}; Galbiati et al., in prep., for full details) allowed us to identify, at high confidence, i3 as another one low-$z$ galaxy interloper. The line emission of i3 in our ALMA band 3 is observed at $109.2\,{\rm GHz}$, the frequency of which possibly corresponds to either the CO(2--1) at $z\simeq 1.11$ or CO(1--0) $z\simeq 0.06$. Future spectroscopic follow-up observations will possibly pin-down the precise redshift of i3. As a result of these analysis, we excluded the aforementioned two sources from the final sample of CO(4--3) line-emitting candidates.

In order to identify other possible additional interlopers belonging to our samples, we inspected band 3 and 6 datacubes in the locations of the CO(4--3), 1.2-, and 3-mm-selected candidates, respectively to look for the presence of any other emission lines. We therefore identified two galaxies: one (dubbed as ``i1'') which belongs to the sample of the 1.2-mm continuum-selected candidates (C06), while the other one (``i2'') which is detected in its 3-mm continuum emission, both showing a very bright emission line in the LSB of the ALMA band 3 datacube. Via the analysis of the MUSE spectra, we accurately determined the redshift of i2 to be $z\simeq 1.42$, therefore identifying the detected line as the CO(2--1). Regarding i1, its redshift determination is highly uncertain given the lack of clear absorption or emission features in the MUSE datacube. However, the carefully inspection of the rest-frame UV spectrum of i1 points to a tentative redshift measurement of $z\simeq 2.54$, thus suggesting that the emission line we detected in our ALMA observations corresponds to the CO(3--2) transition.

\begin{table}[t]
\caption{Redshift bin, cosmological volume, and limiting luminosity of the main emission lines entering in the upper sideband of our ALMA band 3 survey.}  
\label{tbl:interlopers}      
\centering   
\resizebox{\hsize}{!}{
\begin{tabular}{lccc}     
\toprule\toprule
Line$\,^{(1)}$								&Redshift$\,^{(2)}$		&Volume$\,^{(3)}$		&limit $L'$$\,^{(4)}$\\
										&					&(${\rm cMpc^{3}}$)		&($10^{9}\,{\rm K\,km\,s^{-1}\,pc^{2}}$)\\

\cmidrule(lr){1-4}
CO(1--0)									&0.0487--0.0846		&6.9					&0.02\\
CO(2--1)									&1.0973--1.1691		&1183.2				&1.64\\
CO(3--2)									&2.1459--2.2535		&2349.7				&2.47\\
CO(4--3)									&3.1943--3.3379		&3027.7				&2.70\\
CO(5--4)									&4.2426--4.4220		&3414.6				&2.71\\
CO(6--5)									&5.2907--5.5060		&3639.4				&2.63\\
\cmidrule(lr){1-4}
\ci{609}									&3.4775--3.6307		&3879.3				&2.72\\
\bottomrule
\end{tabular}
}
\tablefoot{$^{(1)}$CO and [C\RNum{1}] emission lines entering in our ALMA band 3 observations. $^{(2)}$Redshift range in which the line can be detected. $^{(3)}$Cosmological volume in the redshift range within the effective area of our survey (${\rm PB\ge0.5}$). $^{(4)}$Limiting $5\sigma$ line luminosity assuming ${\rm FWHM} = 300\,{\rm km\,s^{-1}}$.}
\end{table}

We then assessed the nature of our ALMA continuum selected sources at $1.2\,{\rm mm}$ by inspecting their MUSE spectra. Among these, 6 out of 11 (QSO, C01, C02, C06, C08, and C09) have high-confident spectroscopic redshift derived from their rest-frame UV spectra, while 5 out of 11 (C03, C04, C05, and C07) are either not detected in MUSE or the available spectrum does not allow to pin-down a precise spectroscopic redshift. For the first group, the spectroscopic information available places four of them in the proximity of the QSO CTS G18.01 redshift, with three of them (QSO, C01, C02) having CO(4--3) detection, and one (C08) which is only detected in its 1.2-mm dust continuum. The other two sources for which redshift measurements from MUSE are available are C06, and C09. C06 is the 1.2-mm continuum counterpart of the interloper i1, possibly detected via the CO(3--2) in the LSB of our ALMA band 3 observations. The MUSE spectra of C09, instead, exhibits Si\RNum{4} $1394, 1403\,{\rm \AA}$, and ${\rm C\RNum{4}}\,1548,1550\,{\rm \AA}$ absorption lines thus determining that this source is an interloper located at $z\simeq2.874$. At this redshift, we do not expect to detect any bright emission line in our ALMA datacubes. We dubbed this source as ``i5''.

In summary, among all the sources extracted with high fidelity from our ALMA data, we unambiguously identified five interlopers located to different redshifts with respect to the QSO, namely i1 ($z\simeq2.54$), i2 ($z\simeq 1.42$), i3 (either $z\simeq 1.11$ or $z\simeq 0.06$),  i4 ($z\simeq 1.170$), and i5 ($z\simeq 2.874$). Interestingly, these findings are consistent to our predictions based on the CO LFs of blank fields.

However, the subsample of secure sources with two independent redshift measurements from both MUSE and the CO(4--3) line is composed by QSO, L01 (C01), L02 (C02), L04, and L09, all lying within $\pm 1000\,{\rm km\,s^{-1}}$ with respect to the QSO systemic redshift. Regarding the remaining sources, both Object B and L03 (C04) are either revealed in the mm dust continuum with ALMA or have a counterpart in the optical/NIR (see, Appendix~\ref{app:counterparts}). The analysis of VLT/FORS2 UBR colors of L03 (C04) suggests that this source belong to $z\sim3-3.5$ thus supporting the conclusion that L03 is actually detected via CO(4--3) at $z\simeq 3.25$. Finally, the recently acquired spectrum of the Object B with the NIR spectrograph on board of JWST definitely confirms that this source is located in the proximity of the QSO (Pensabene et al., in prep.). On the other hand, L05, L06, L07, and L08 are only detected in line with ALMA. Therefore, we cannot rule out that the latters sources actually represent either false-positive detections or interlopers located at different redshift. Future deep NIR observations are needed to assess the nature of such objects. Interestingly, these sources exhibits a large velocity shift and significant spatial separation from the QSO. In what follows, we took the aforementioned consideration into account in the computation of the CO LF.

\subsection{Source fluxes and luminosities}\label{ssect:flux_measure}
The majority of CO(4--3) and continuum emitters detected in this work appear as compact spatially-unresolved sources. For such objects, the total continuum or line flux can be measured through a standard single-pixel analysis of the data. However, in the case of partially-resolved objects or extended sources with complex morphology, this simple method yields to significant flux underestimation. In this work, for such sources we therefore performed source flux measurements by applying the $2\sigma$-clipped photometry\footnote{This approach has been demonstrated to yield consistent results with respect to other commonly-used methods such as 2D fit of the flux map or aperture photometry.} \citep[see, e.g.,][]{Bethermin+2020} as described below.

We measured the 1.2- {\rm and 3-mm} flux of the sources from the cleaned ALMA band 6 {\rm and 3} continuum images, {\rm respectively}, not corrected for the combined mosaic PB response. For each source, we sum the flux density in ${\rm mJy\,beam^{-1}}$ enclosed in the contiguous area around the source including all pixels with ${\rm S/N\ge2}$\footnote{We measured the noise as the standard deviation of the signal on the whole image.}. We then divided the total flux density per beam by the synthesized beam area (in pixel units) and we rescaled the flux for the PB response at the location of the source. We computed the flux uncertainty by rescaling the noise by the square root of the number of independent beams within the integration area. In Table~\ref{tbl:1.2mm_cand} we report our source flux measurements as well as their peak flux in the continuum. To understand which sources can be considered point-like, we computed the uncertainty-normalized difference between the two flux estimates as $\Delta_{F}=(F_{2\sigma}-F_{\rm peak})/\sqrt{\sigma_{2\sigma}^{2} + \sigma_{\rm peak}^{2}}$, where $F_{2\sigma}$, $F_{\rm peak}$, $\sigma_{2\sigma}$, and $\sigma_{\rm peak}$ are the 1.2-mm continuum flux obtained via the $2\sigma$-clipped photometry, the peak flux, and their uncertainties, respectively. Within this formalism, we expect sources which are significantly resolved to have $\Delta_{F} > 1$. Sources C01 and C02 are such cases; for them we therefore adopted $S_{\rm 1.2\,mm}^{2\sigma}$ as our fiducial 1.2-mm continuum flux density measurement\footnote{We additionally verified our results by performing 2D fit of continuum images by using the CASA task {\tt imfit}.}. We however note that the proper flux measurement of QSO CTS G18.01, and its closely-separated companion (Object B) is challenging due to the partial blending of the sources at the current resolution of the continuum data (see Sect.~\ref{ssect:data_reduction}). In order to minimize the flux contamination, for such sources we adopted their 1.2-mm continuum flux peak $F^{\rm peak}_{\rm 1.2\,mm}$.

We measured the CO line fluxes by following the iterative process presented in \citet{Bethermin+2020}. As for the continuum, the CO line emission of the QSO CTS G18.01 and Object B are partially blended. To measure their fluxes in what follows, we employed the ALMA band 3 high-resolution ($\sim 0.25\arcsec$) data (see, Sect.~\ref{ssect:ancillary_data}) where the line emission of the sources are spatially-resolved. For each CO(4--3)-line emitter we extracted the spectrum at the peak position of the source from the ``cleaned'' ALMA band 3 datacube binned at $40\,{\rm km\,s^{-1}}$. We then scaled the source spectrum by the PB response, and we performed a fit using a Gaussian profile for the line and a constant for the underlying 3-mm dust continuum using the \texttt{curve$\textunderscore$fit} task included in the SciPy package \citep{Scipy}. We then produced the line-velocity integrated map using all the channels within $\pm2\sigma$ from the line centroid. Subsequently, we re-extracted the source spectrum by summing all the spectra in pixels showing ${\rm S/N \ge 2}$ in the moment-0 map, and we rescaled the channel fluxes and their uncertainties respectively by the synthesized beam area, and the square root of the number of independent beams within the integration area. In this process, we masked all the pixels {\rm below} the chosen threshold which we confidently believe to be not related to any real emission from the source. This new spectrum is more informative in the case of (partially-)resolved sources since it includes the signal from the entire source line-emitting region. We therefore performed a new spectral fit using the best-fit parameters of the previous iteration as starting point for the fitting code. We repeated the aforementioned steps a few times until convergence. All the extracted source spectra remain stable after a few ($<10$) iterations. Similarly to continuum flux estimates discussed above, we computed the quantity $\Delta_{F}$ for each source. As a result, L01, L02, the QSO, and the Object B\footnote{The CO emission line of the QSO and Object B is spatially-resolved in the high-resolution observations. For them automatically follows $\Delta_{F} > 1$.} all exhibit $\Delta_{F} > 1$. Accordingly, this analysis determined our final source spectra. We finally performed a finer fit of the final source spectrum by sampling the parameter space through the Python Monte Carlo Markow Chain (MCMC) ensemble sampler \texttt{emcee} \citep{Foreman+2013}. We employed flat priors on the basis of the best-fit parameters derived from the last iteration and we assume Gaussian uncertainties in the definition of the likelihood. We finally derived the line luminosities as \citep[see, e.g.,][]{Solomon+1997}:
\begin{align}
\label{eq:lum}
&L_{\rm CO}\quadra{L_{\astrosun}}=1.04\times10^{-3}S\Delta\varv\,\nu_{\rm obs}\,D_{L}^{2},\\
&L'_{\rm CO}\quadra{{\rm K\,km\,s^{-1}\,pc^{2}}}=3.25\times10^{7}S\Delta\varv\frac{D_{L}^{2}}{\tonda{1+z}^{3}\nu_{\rm obs}^{2}},
\label{eq:lum1}
\end{align}
where $S\Delta\varv$ is the velocity-integrated line flux in ${\rm Jy\,km\,s^{-1}}$, $\nu_{\rm obs}$ is the observed central frequency of the line in GHz, $z$ is the source redshift measured from the CO line centroid, and $D_{L}$ is the luminosity distance in Mpc. The relation between Eq.~(\ref{eq:lum}) and~(\ref{eq:lum1}) is $L_{\rm CO}=3\times10^{-11}\,\nu_{\rm rest}^{3}L'_{\rm CO}$, where $\nu_{\rm rest}$ is the line rest frequency in GHz. 

   \begin{figure*}[!t]
   	\centering
   	\resizebox{0.8\hsize}{!}{
      		\includegraphics{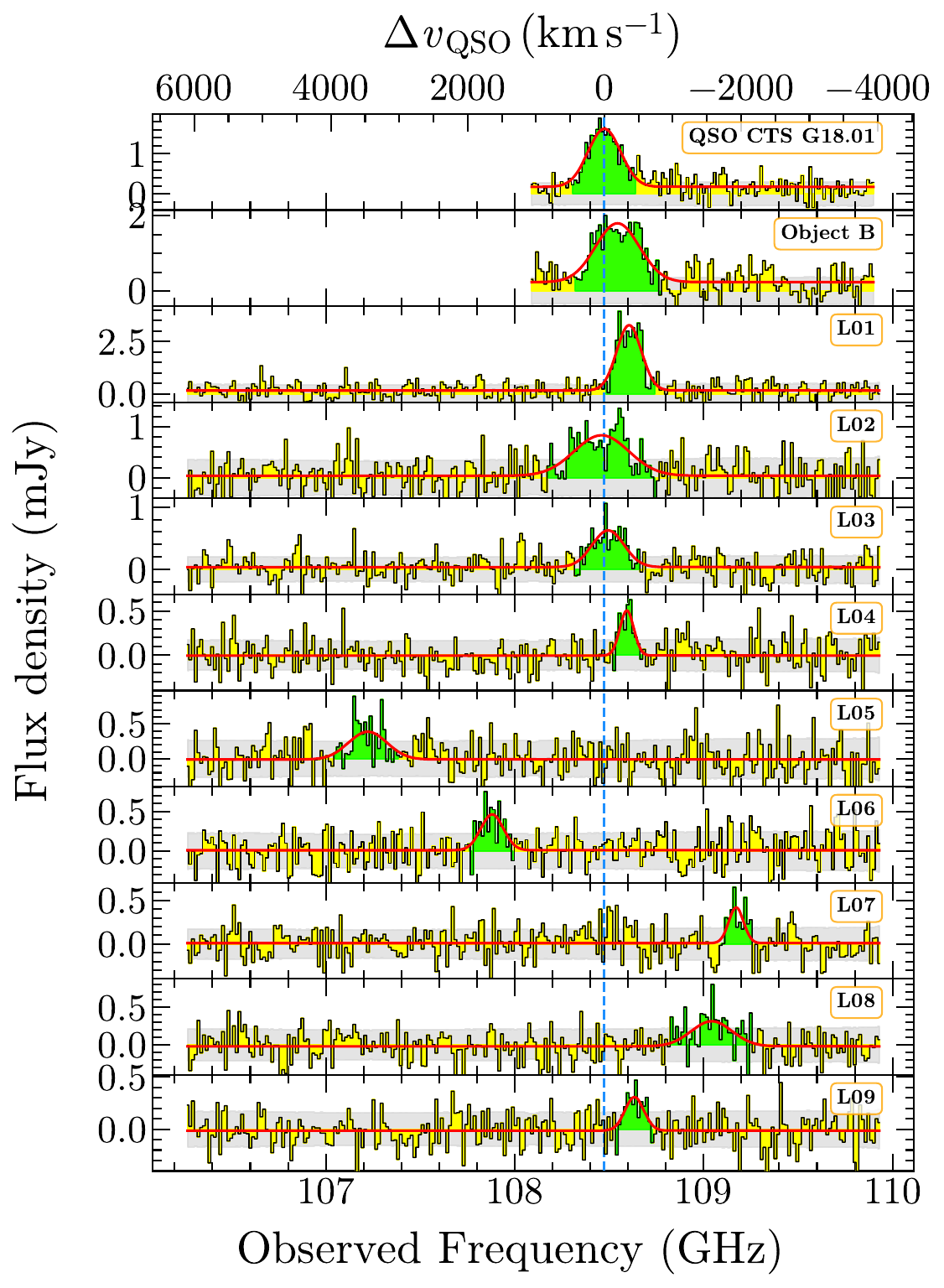}}
       \caption{Spectra of the CO(4--3) line-emitting sources in our ALMA survey of the MQN01 field. The red lines are the best fit models to the spectra (yellow bins). The green bins indicate the channels we used to compute the line-velocity integrated maps (see Fig.~\ref{fig:co_maps}) defined within $\pm 2\sigma$ of the best-fitting Gaussian line. The horizontal gray bands are the rms noise in each channel. The blue vertical line indicate the QSO CTS G18.01 systemic velocity. Sources are labeled according to the ID$_{\rm CO}$ reported in Table~\ref{tbl:CO_cand}.}
       \label{fig:co_spectra}
   \end{figure*}
   
   \begin{figure*}[!t]
   	\centering
   	\resizebox{0.8\hsize}{!}{
      		\includegraphics[angle=0]{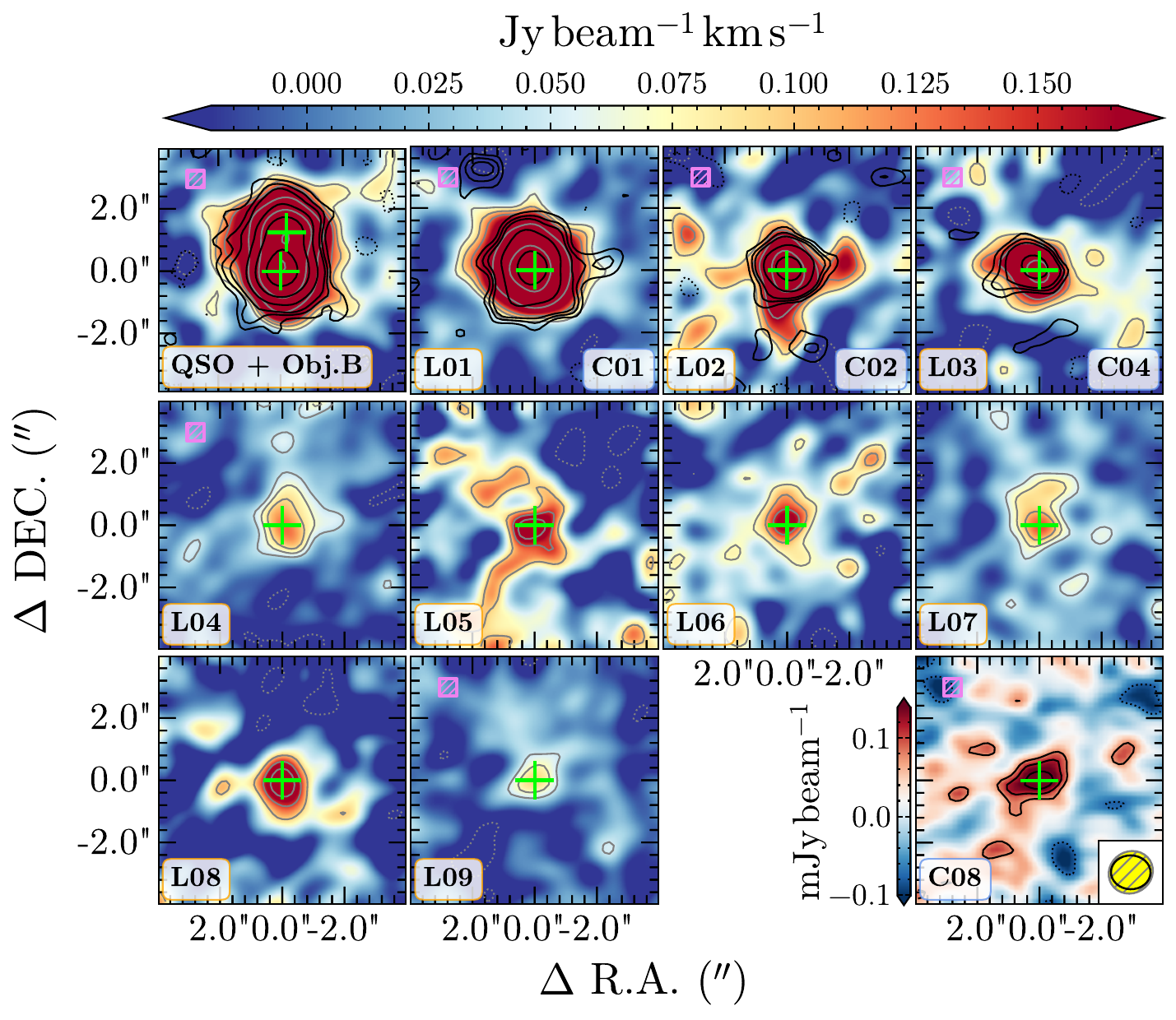}}
       \caption{Maps of the ALMA-selected galaxies in MQN01 field either detected in their CO(4--3) line or in continuum at 1.2-mm, or both. The color scale at the top refers to line-velocity integrated maps for all sources except C08, which is only detected at 1.2-mm in the dust continuum (bottom right color scale). Grey and black solid contours correspond to the velocity-integrated and 1.2-mm dust continuum maps, respectively and scales as $[2,3,2^{N}]\times\sigma$ with $N>1$ integer number. Dotted contours are the $-2\sigma$ level. The yellow ellipse drawn in the bottom right corner represents the FWHM of the synthesized beam of the ALMA band 3 (yellow fill and gray line), and 6 (black line) observations. Sources are labeled according to the ID$_{\rm CO}$ and/or ID$_{\rm 1.2\,mm}$ reported in Table~\ref{tbl:CO_cand}. Sources with spectroscopic-redshift confirmation are marked with a violet square in the upper-left corner.}
       \label{fig:co_maps}
   \end{figure*}
   
As our observations probe the (rest-frame) FIR wavelengths of high-$z$ galaxies, the Cosmic Microwave Background (CMB) might have an impact on our continuum and line measurements by increasing both the dust and the line excitation temperature. In addition, the CMB provides a strong background signal \citep[see,][]{DaCunha+2013, Zhang+2016}. 

By assuming a modified black body with typical values for SMGs at $z\sim 3$ \citep[i.e., spectral index 1.6, and dust temperature of $35\,{\rm K}$; see, e.g.,][]{Kovacs+2006}, the CMB affects our  $1.2\,{\rm mm}$ dust continuum measurements by $\lesssim 5\%$, thus compatible with the typical uncertainties on our flux estimates. The CMB effect further decreases assuming higher dust temperature.

Regarding the CO(4--3) line measurements, under the assumption of local thermodynamic equilibrium (LTE; i.e., the line excitation temperature equaling the gas kinetic temperature), the effect of CMB in reducing the recovered line fluxes is always $<20\%$ for any excitation temperature $> 40\,{\rm K}$ at $z\lesssim 4$ \cite[see,][]{Decarli+2020}. However, due to the unknown excitation temperature of the gas and the unverified assumption of LTE for our sources, in this work we opted to not apply the CMB-related corrections \citep[see, e.g.,][]{Decarli+2020, Boogaard+2023}.
   
We list the measured fluxes and derived quantities of the lines in Table~\ref{tbl:CO_cand}, and we report the final source spectra in Fig.~\ref{fig:co_spectra}. In Fig~\ref{fig:co_maps} we report the source maps of the combined sample sources within $\abs{\Delta\varv_{\rm QSO}}<4000\,{\rm km\,s^{-1}}$ either detected in their CO(4--3) line or continuum at 1.2 mm.

 \section{Results}\label{sect:results}
 
 \subsection{CO Luminosity Function Analysis}\label{ssect:co_lf}
We computed the CO(4--3) LF by following the approach described in \citet{Decarli+2016, Decarli+2019c, Decarli+2020, Riechers+2019, Boogaard+2023}. We define the CO LF as
 \eq{\Phi_{\rm CO}\,\tonda{{\rm log}\,L'_{\rm CO}} \quadra{\rm cMpc^{-3}\,dex^{-1}} = \frac{1}{\Delta V\,\Delta\,{\rm log}\,L'_{\rm CO}}\sum_{i}\frac{F_{i}}{C_{i}},}
where $\Phi$ is the number of sources per comoving ${\rm Mpc^{3}}$ in the luminosity interval ${\rm log}\,L'_{\rm CO}\pm 0.5\,\Delta({\rm log}\,L'_{\rm CO})$, $\Delta V$ is the comoving volume of our survey, and $F_{i}$ and $C_{i}$ are the fidelity and completeness associated to each source, respectively. We computed two different CO LFs, one in $\Delta V$ corresponding to the redshift range within $\pm 4000\,{\rm km\,s^{-1}}$ with respect to the systemic redshift of QSO CTS G18.01 ($\Delta\varv_{\rm QSO}$), and another one within $\abs{\Delta\varv_{\rm QSO}} < 1000\,{\rm km\,s^{-1}}$. The corresponding cosmological volumes are respectively $\Delta V_{4000}=2395\,{\rm cMpc^{3}}$, and $\Delta V_{1000}=599\,{\rm cMpc^{3}}$. While $\Delta V_{4000}$ contains all our eleven CO(4--3) line-emitting candidates, $\Delta V_{1000}$ encompasses only sources having a optical/NIR counterparts (see also, Appendix~\ref{app:counterparts}) with a spectroscopic-confirmed redshift, thus including secure sources ($F=1$). In this regard, we use sources in $\Delta V_{1000}$ to obtain a ``raw'' CO LF without applying the completeness correction. 

\begin{figure*}[!t]
   	\centering
   	\resizebox{\hsize}{!}{
		 \includegraphics{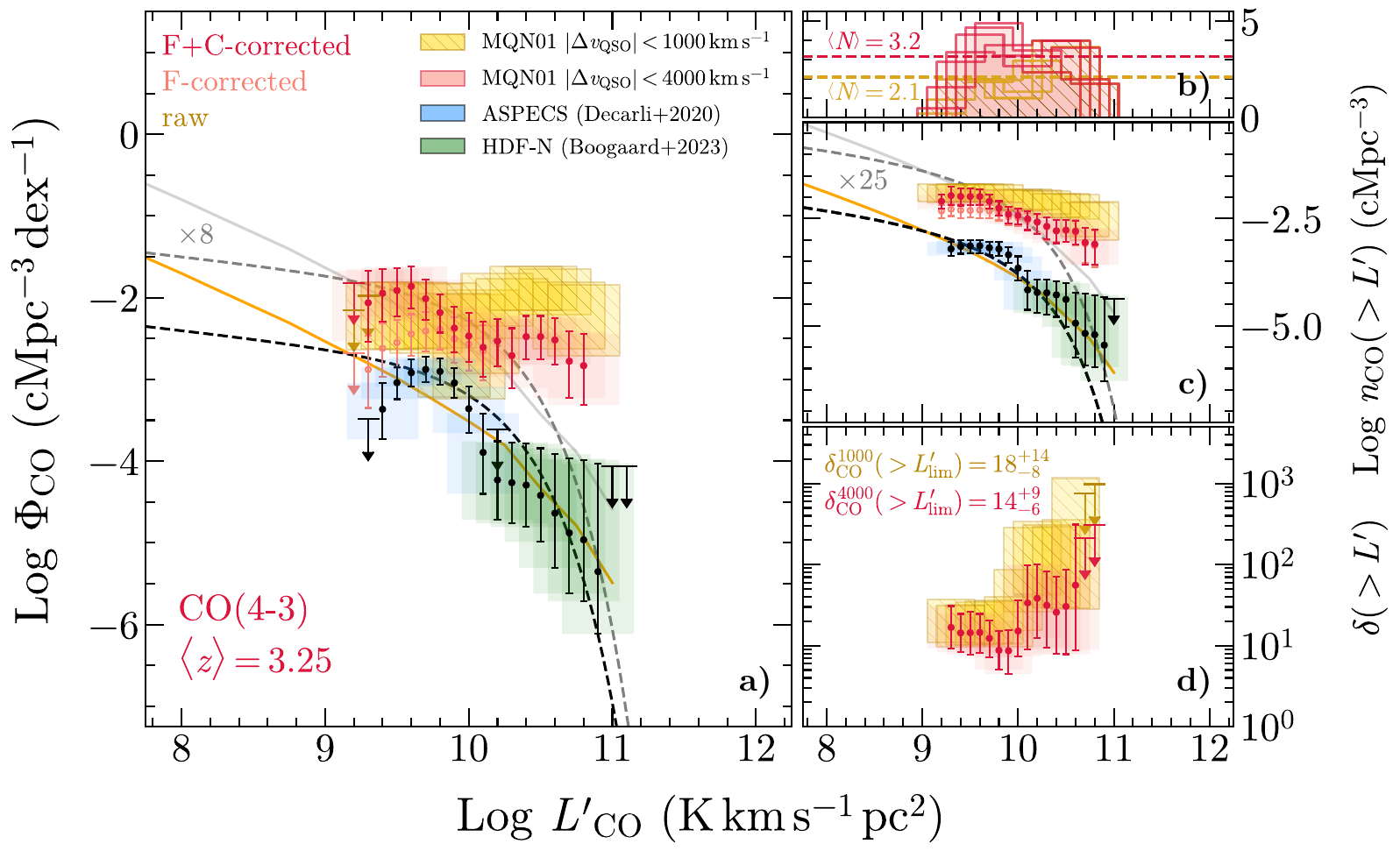}}
       \caption{The overdensity of CO emitters in the MQN01 field. {\it Panel a):} CO(4--3) LF in MQN01 and in blank fields. The (dark/light) red dots and boxes represent the CO LF in MQN01 field (fidelity (F) + completeness (C)-corrected/uncorrected, respectively) in $\Delta V_{4000}$, while gold hatched boxes are the uncorrected (raw) CO LF within $\Delta V_{1000}$. The blue and green boxes with black dots are data from ASPECS and HDF-N, respectively, which are representative of blank fields at these redshifts \citep[e.g.,][]{Decarli+2019c, Decarli+2020, Boogaard+2023}. The downward arrows indicate 1-$\sigma$ upper limits. The dashed black line is the best-fit Schechter function to the ASPECS+HDF-N data reported in \citet{Boogaard+2023}. The orange line is the CO(4--3) LF prediction from SIDES simulation \citep{Bethermin+2022, Gkogkou+2023} at $z=3-3.25$. Grey lines represent the rescaled versions of such models. {\it Panel b)}: Number of sources in each luminosity bin averaged over the iterations. The horizontal dashed line reports the total average number of sources entering in each bin of the LFs. {\it Panel c):} Cumulative source number density as a function of the CO(4--3) line luminosity obtained by integrating the CO LFs. {\it Panel d):} Source overdensity in MQN01 field as a function of the CO(4--3) line luminosity obtained by a bin-to-bin ratio of the cumulative source number densities.}
       \label{fig:CO_LF}
   \end{figure*}

To obtain the CO LF in the MQN01 field, we performed a Monte Carlo simulation of 1000 independent realizations of the LF. In each iteration, we varied the CO line luminosity, line FHWM, and the source fidelity of our CO line emitters within their uncertainties. For secure sources, we fixed the fidelity value to $F=1$, for all the other sources we treated the fidelity as upper limit. {\rm This approach provides a conservative treatment of our fidelity estimates attempting to include the systematic uncertainties associated with sources without any clear multiwavelength counterparts \citep[see, e.g.,][]{Pavesi+2018, Decarli+2019c, Riechers+2019}.} In each iteration, we extracted a number ($P$) from a random uniform distribution in the interval $[0;1]$. We then selected those sources entering in the realization having $P\le F$. In this way, sources with higher fidelity have a larger chance to be selected. We computed the number of sources and associated $1\sigma$ Poisson confidence intervals \citep{Gehrels1986} in $0.5\,{\rm dex}$ bins. We then rescaled the resulting counts and uncertainties by the completeness corrections\footnote{As previously mentioned, we did not apply the completeness correction for the CO LF in $\Delta V_{1000}$.}, and then divided them by the effective volume of the survey and by the luminosity bin width. Finally, we averaged over the realizations. We repeated the entire simulation five times by shifting the luminosity bins of $0.1\,{\rm dex}$ in order to mitigate the dependence of the result on the bin definitions and to expose the intra-bin variations. For bins with a less than one count on average, we report a $1\sigma$ upper limit. The resulting CO(4-3) LF is shown in Fig.~\ref{fig:CO_LF} (panel {\it a}; red and gold symbols, see caption for a detailed explanation). In the same figure, we also show the LF derived from blank fields at the same redshift reported from the literature. 

Our aim is to compare the CO LF in MQN01 with that of blank fields. To do so, we combined data from ASPECS \citep[The ALMA Spectroscopic Survey in the Hubble Ultra Deep Field][]{Walter+2016, Gonzalez-Lopez+2019, Decarli+2019c, Decarli+2020}, and \citet{Boogaard+2023} who performed a blind search of molecular lines in the Hubble Deep Field North (HDF-N) using the Northern Extended Millimeter Array (NOEMA). These works provide us with the most up-to-date CO(4--3) LF at $z\simeq3.5$. To enable a direct comparison with our result in the MQN01 field, we recomputed LFs consistently as described above adopting common luminosity bins. 
Additionally, we obtained the CO(4--3) LF from the SIDES simulation \citep[Simulated Infrared Dusty Extragalactic Sky;][]{Bethermin+2022, Gkogkou+2023} at $z=3-3.5$. As a comparison, in Fig.~\ref{fig:CO_LF}, we also rescaled by a factor of $8\times$ the best-fit Schechter function to the blank fields \citep[see,][]{Boogaard+2023} as well as the CO LF predicted from SIDES. Interestingly, the CO LF in MQN01 differs significantly not only in normalization but also in shape with respect to that of blank fields showing a flattening at its bright end.  We further discuss this point in Sect.~\ref{sect:discussion}. 

We then computed the cumulative source number counts per comoving volume $n_{\rm CO}(>L'_{\rm CO})$ by integrating the five different CO LFs separately, each of which has uncorrelated luminosity bins. We show our results in Fig.~\ref{fig:CO_LF} ({\rm panel {\it c}}). We also obtained cumulative source number counts from both the best-fit of blank fields and SIDES by integrating the corresponding CO LFs. In Fig.~\ref{fig:CO_LF}, we also rescaled the latter functions by a factor of $25\times$ as a comparison to our results.

Finally, we evaluated the galaxy overdensity $\delta(>L'_{\rm CO})$ as traced by CO(4--3) by computing the bin-to-bin ratios between the cumulative number counts in MQNQ01 and those in blank fields. The measured cumulative overdensity of CO(4--3) line emitters is a strong function of luminosity, increasing from $\sim8$ (at the lowest luminosities) to $\sim 200$ (at the highest luminosities). Above the limiting CO luminosity of our survey we estimated $\delta_{\rm CO}^{4000}(>L'_{\rm lim})= 14^{+9}_{-6}$, and $\delta_{\rm CO}^{1000}(>L'_{\rm lim})= 18^{+14}_{-8}$, in $\Delta V_{4000}$, and $\Delta V_{1000}$, respectively. The overdensity plot is also shown in Fig.~\ref{fig:CO_LF} (panel {\it d}).

\subsection{1.2-mm continuum source number counts}\label{ssect:12mm_counts}
We investigate the galaxy overdensity in the MQN01 field as traced by dust continuum emission at $1.2\,{\rm mm}$. Thanks to the combination of our deep spectroscopic VLT/MUSE and ALMA surveys we can pin-down the redshift of a sufficiently large sample of continuum-selected galaxies in the MQN01 field. We computed the source count density of our continuum-selected galaxies in a given cosmological volume around the CTS QSO G18.01 quasar by adapting the recipe from (e.g.,) \citet[][]{Hatsukade+2013, Hatsukade+2016, Hatsukade+2018, Carniani+2015, Aravena+2016, Fujimoto+2016, Fujimoto+2023, Umehata+2017,Umehata+2018, Gonzalez-Lopez+2019, Gonzalez-Lopez+2020}. We defined the differential source number count density per flux bin $S_{1.2\,{\rm mm}}$ as
\eq{\frac{dn}{d({\rm log S})} \quadra{\rm cMpc^{-3}\,dex^{-1}} = \frac{1}{\Delta V\,\Delta\,{\rm log}\,S}\sum_{i}\frac{F_{i}}{C_{i}},}
or,
\eq{\frac{dn}{dS} \quadra{\rm cMpc^{-3}\,mJy^{-1}} = \frac{1}{\Delta V\,\Delta\,{\rm log\,S}\,\ln(10)\,S}\sum_{i}\frac{F_{i}}{C_{i}}.\label{eq:diff_counts}}

In the computation of the source number density, we included all our $1.2\,{\rm mm}$-continuum selected sources having spectroscopic redshift information either from the analysis of VLT/MUSE spectrum, or from their CO(4--3) line which lie within $\abs{\Delta\varv_{\rm QSO}}<1000\,{\rm km\,s^{-1}}$ ($\Delta V=430\,{\rm cMpc^{3}}$)\footnote{This volume slightly differs from $\Delta V_{1000}$ used in the computation of the CO LF because of the different survey sky area of the ALMA band 6 mosaic from which we extracted the 1.2-mm continuum-emitting candidates.}. This selection yielded to a sample of six sources (including the QSO and the Object B; see Table~\ref{tbl:1.2mm_cand}, and~\ref{tbl:CO_cand}). 

To compare our results to blank fields, we employed the ASPECS publicly-available catalogs\footnote{\url{https://aspecs.info/data}} \citep{Aravena+2019, Aravena+2020, Boogaard+2019, Boogaard+2020, Gonzalez-Lopez+2019, Gonzalez-Lopez+2020} of the ALMA band 6 survey from which we select all $1.2\,{\rm mm}$-continuum detected galaxies with spectroscopic redshift information. To increase the statistics of the sample we considered all sources with $2.5\leq z\leq 3.5$. This selection effectively yielded to a sample of ten galaxies in the redshift range $z=2.543-2.981$ (median redshift $z=2.685$). Conservatively, we employed such range to compute the corresponding cosmological volume.

    \begin{figure*}[!t]
   	\centering
   	\resizebox{\hsize}{!}{
      		\includegraphics{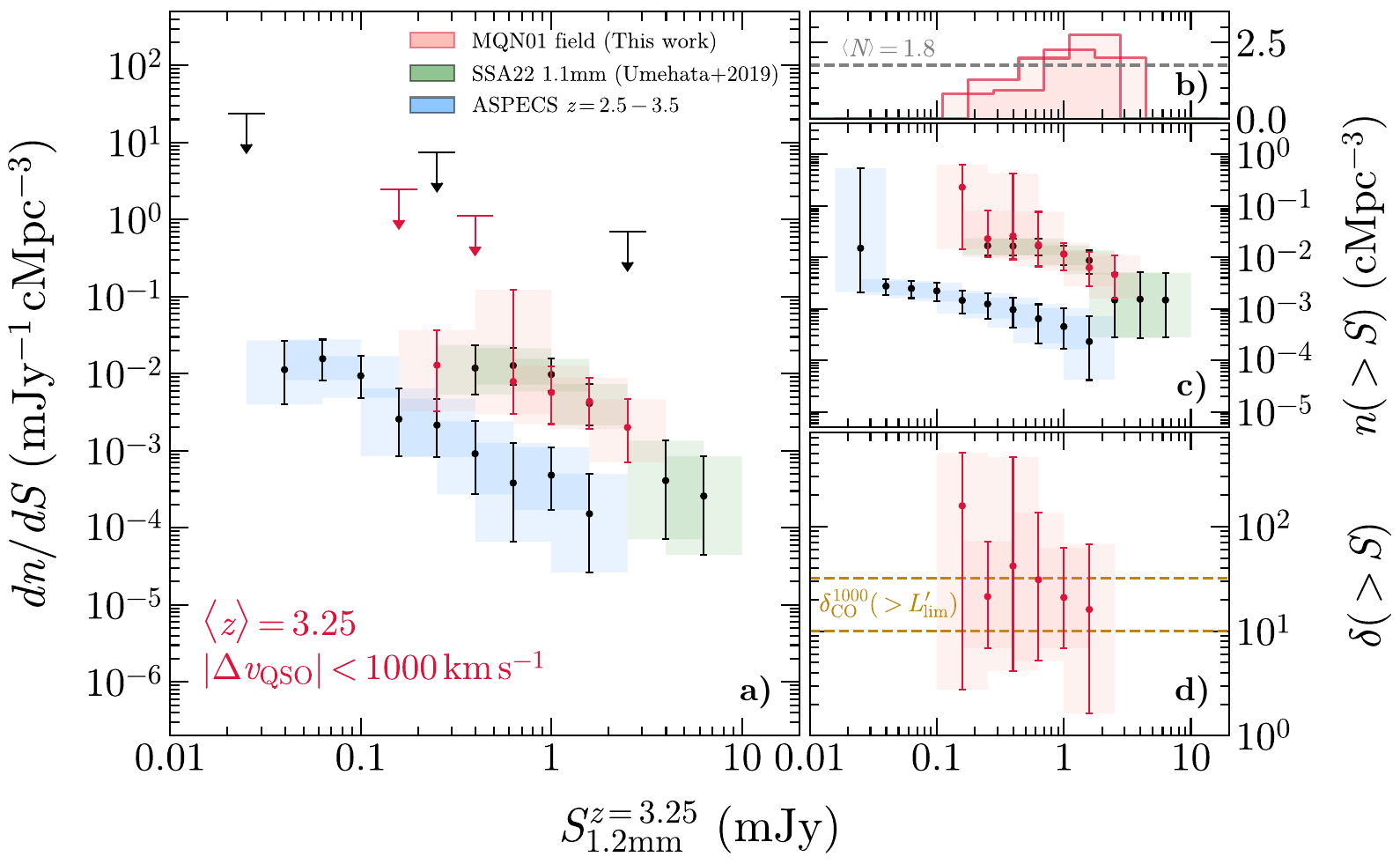}}
       \caption{The overdensity of dusty star forming galaxies in the MQN01. {\it Panel a):} Differential $1.2\,{\rm mm}$ continuum source number count density in MQN01 (red points and boxes; including galaxies within $\abs{\Delta\varv_{\rm QSO}}<1000\,{\rm km\,s^{-1}}$); SSA22 \citep[black dots and green boxes;][within $\pm1000\,{\rm km\,s^{-1}}$ with respect to the median source redshift]{Umehata+2017, Umehata+2019}, and in blank fields as derived from ASPECS large program \citep[black points and blue boxes;][]{Aravena+2019, Aravena+2020, Boogaard+2019, Boogaard+2020, Gonzalez-Lopez+2019, Gonzalez-Lopez+2020} including sources with spectroscopic redshift between $2.5\le z \le 3.5$. The downward arrows indicate 1-$\sigma$ upper limits. {\it Panel b)}: Number of sources in each luminosity bin averaged over the iterations. The horizontal gray line reports the total average number of sources entering in each bin for the computation of the differential number counts. {\it Panel c):} Cumulative source number density as function of $1.2\,{\rm mm}$ continuum flux obtained by integrating the differential number count density. {\it Panel d):} Cumulative source overdensity in MQN01 as a function of $S_{1.2\,\rm mm}$. The horizontal lines represent the overdensity range of CO(4--3) emitters in $\Delta V_{1000}$.}
       \label{fig:12mm_LF}
   \end{figure*}
   
Additionally, we compared our source number count density with that observed toward the SSA22 field which is found to show a large overdensity of Ly-$\alpha$ emitters and sub-millimeter galaxies (SMGs) at $z\sim 3$ \citep[see,][]{Steidel+1998, Steidel+2000, Yamada+2012, Umehata+2015, Umehata+2017, Umehata+2018, Umehata+2019}. To this purpose, we cross-matched the catalog of the CO(3--2) emitters with that of continuum-selected galaxies at $1.1\,{\rm mm}$ as revealed by the ALMA deep field survey in the SSA22 field A \citep[ADF22A; see,][]{Umehata+2017, Umehata+2019}. The final sample comprises ten sources with redshift measurement within $\pm 1000\,{\rm km\,s^{-1}}$ around the median redshift of the sources $z=3.0951$.

The sources we selected from ASPECS surveys are observed at $1.2\,{\rm mm}$ but lie in a different redshift range, while that from \citet{Umehata+2017, Umehata+2019} are observed at $1.1\,{\rm mm}$. In order to mitigate possible biases introduced in the selection of sources at different redshifts and which are observed at different wavelengths, we employed a ``$k$-correction'' to translate the observed monochromatic flux $S_{\!\lambda\rm obs}(z)$ of a source at redshift $z$ to $S_{\rm 1.2\,mm}(z_{0})$ defined as 

\begin{align}&k_{\lambda\rm obs}(z)=\frac{S_{\rm 1.2\,mm}(z_{0})}{S_{\!\lambda \rm obs}(z)}=\\
&=\frac{1+z_{0}}{1+z}\quadra{\frac{D_{L}(z)}{D_{L}(z_{0})}}^{2}\tonda{\frac{\nu_{0}}{\nu}}^{\beta}\frac{B_{\nu_{0}}[T_{\rm dust}(z_{0})]-B_{\nu_{0}}[T_{\rm CMB}(z_{0})]}{B_{\nu}[T_{\rm dust}(z)]-B_{\nu}[T_{\rm CMB}(z)]\nonumber},
\end{align}

where $\nu_{0}=c(1+z_{0})/\lambda_{1.2\,{\rm mm}}$ and $\nu=c(1+z)/\lambda_{\rm obs}$. Here we estimated $k_{\rm \lambda obs}$ by assuming a modified black body emission with typical dust temperature ranging in $T_{\rm dust}=20-45\,{\rm K}$, and spectral index $\beta=1.5-2.0$. We additionally take into account the contrast effect produced by the CMB at high-$z$ \citep[see,][]{DaCunha+2013} the temperature of which ranges between $T_{\rm CMB}=9.5-12.3\,{\rm K}$ at $z=2.5-3.5$. Under such assumptions and by using a reference redshift of $z_{0}=3.25$, we found $k_{\rm 1.2\,mm}(z =2.5) =1.11^{+0.06}_{-0.10}$, $k_{\rm 1.2\,mm}(z=3.5) = 0.97^{+0.03}_{-0.02}$, and $k_{\rm 1.1\,mm}(z=3.1) = 0.796\pm 0.009$, the uncertainties of which we took into account in converting the observed fluxes and computing the source number counts. 

We therefore computed the differential and cumulative source number count densities, as well as source overdensity in MQN01, by adopting the same approach described in Sect.~\ref{ssect:co_lf}. We used common bins for all the different fields corresponding to $\Delta\,{\rm log}\,S=0.4$, and we repeated the simulation two times by shifting the flux bins of $0.1\,{\rm dex}$. Since all the sources involved in this computation are detected via multiple tracers, we set the fidelity of all sources to $1$. In addition, we opted to ignore the completeness corrections which we expect to be not relevant due to the large uncertainties involved in such analysis. We report our results in Fig.~\ref{fig:12mm_LF}. Overall, the 1.2-mm source number count density in MQN01 shows higher normalization and similar shape with respect that of blank fields at similar redshift without any evidence of a flattening at the bright end. Interestingly, our result resembles the source number count density measured in SSA22. Overall, the overdensity of sources detected in continuum at ${\rm 1.2\,mm}$ in the MQN01 field within $\abs{\Delta\varv_{\rm QSO}} < 1000\,{\rm km\,s^{-1}}$ is consistent with that estimated in the same redshift range by using CO emitters. We further discuss this point in Sect.~\ref{sect:discussion}.

\section{Discussion}\label{sect:discussion}

\subsection{The overdensity in MQN01}
In Sect.~\ref{ssect:co_lf} we obtained the CO LF in the MQN01 field within $\abs{\Delta \varv_{\rm QSO}} < 4000\,{\rm km\,s^{-1}}$, and $\abs{\Delta \varv_{\rm QSO}} < 1000\,{\rm km\,s^{-1}}$. The analysis we performed on the data allowed us to estimate the presence of a galaxy overdensity traced by the molecular gas content. We found clear evidence of an overdensity in MQN01 which is a factor of $\sim$10 -- 100 higher than the field, depending on the luminosity cut. The luminosity dependence of the overdensity indicates that the most luminous objects are also the most overabundant. Taking into account all the CO line-detected sources (above the limiting CO(4--3) luminosity $L'_{\rm lim}$), we estimated $\delta_{\rm CO}^{4000}(>L'_{\rm lim})=14^{+9}_{-6}$ in $\Delta V_{4000}$, and $\delta_{\rm CO}^{1000}(>L'_{\rm lim})=18^{+14}_{-8}$ counting only secure sources within $\Delta V_{1000}$ without applying the completeness correction. Our results suggest that galaxies in MQN01 field are possibly part of a structure extending within a cosmological volume of at least $\sim 600\,{\rm cMpc^{3}}$ as probed by our ALMA survey. 

In addition, we can investigate the normalization and shape of the CO LF in MQN01 which encloses key information on the assembly of galaxies in such an overdense region. However, an accurate analysis of the LF trend and its interpretation is challenging given the low statistics of the sample and large uncertainties. To compare the CO LF in MQN01 with that measured in blank fields, we considered the one computed in $\Delta V_{4000}$ corrected for both the source fidelity and the survey completeness.
   
The observed CO LF in the MQN01 field can be produced by a combination of an excess in galaxy number counts (thus increasing the overall normalization), and an enhanced CO luminosity of galaxies in the field (which translates to a shift of the source counts toward higher luminosities). However, a pure luminosity shift seems at odds with the data. Indeed, assuming that the CO LF in MQN01 follows the trend expected for blank fields (typically fit by a Schechter function; \citealt{Schechter1976}), there is evidence for a further excess at the high-luminosity end producing a flattening at $\log L'_{\rm CO}/({\rm K\,km\,s^{-1}\,pc^{2}}) \apprge 10$, which is also reflected in an increase of the cumulative overdensity values $\delta(>L'_{\rm CO})$ reported in Fig.~\ref{fig:CO_LF}. Interestingly, this excess is even more evident when considering the ``raw'' CO LF within $\Delta V_{1000}$. To demonstrate this fact, in Fig.~\ref{fig:CO_LF} we report the rescaled version of the best-fit model to blank field data and the result obtained from the SIDES simulation. If the observed flattening at the high-luminosity tail of the LF is ascribed to a systematical increase of galaxy CO(4--3) luminosities, this can be either due to a higher galaxy molecular gas budget, or enhanced molecular gas excitation due to a significant star-formation activity and/or higher AGN fraction in the field which can increase the fraction of molecules populating higher-$J$ CO states without necessary increasing the bulk of the CO-traced molecular gas mass. Interestingly, four out of seven ($\simeq 57\%$) of the CO-detected sources in $\abs{\Delta\varv_{\rm QSO}}< 1000\,{\rm km\,s^{-1}}$ are also detected in the X-rays thus suggesting an intense AGN activity in the field (Travascio et al., in prep.). In the first scenario, if such galaxies follow the Kennicutt-Schmidt (KS) relation \citep[][for a review]{Schmidt1959, Kennicutt1998, Kennicutt+2012}, this would imply that galaxy assembly is accelerated in this dense field with more massive galaxies experiencing considerable episodes of star-formation. Regarding the second possible scenario, the sole detection of CO(4--3) is not sufficient to determine the dominant physical mechanism responsible for the molecular gas excitation. A proper sampling of multiple CO rotational ladders is therefore needed to understand if there is an important contribution to the enhancement of the CO(4--3) luminosity from high star-formation activity or X-ray radiation from AGN \citep[see, e.g.,][for a review]{Wolfire+2022}. Indeed, highly excited molecular gas and high star-formation efficiency have been found in the core of protoclusters at high-$z$ \citep[see, e.g.,][]{Lee+2017, Coogan+2018}. However, since the aforementioned effects are interlaced, we are hampered in drawing strong conclusions here. Further information can be obtained by evaluating the SFR of the selected CO line-emitting galaxies, which is possible in the mm regime by measuring the galaxy FIR luminosities \citep[see, e.g.,][]{DeLooze+2014}. However, our observations provide us with at best two different photometric measurements in the Rayleigh-Jeans (RJ) tail of the dust SED, therefore preventing a proper estimation of the $L_{\rm FIR}$ without an a-priori assumption on the galaxy dust temperature ($T_{\rm dust}$).

The comparison of the CO LFs is based on the evaluation of differential number counts which are likely to be affected by Poissonian statistical oscillations. Cumulative luminosity function $n_{\rm CO}(>L')$ can mitigate the uncertainties providing a more robust result. Here, the flattening that occurs at $\log L'_{\rm CO}/({\rm K\,km\,s^{-1}\,pc^{2}}) \apprge 10$ is less sharp. However, there is a hint of a shallower decline of $n_{\rm CO}$ with respect to blank fields.

We can then ask if the overdensity revealed in MQN01 via CO line emission is different from that revealed by other tracers. In Sect.~\ref{ssect:12mm_counts}, we analyzed the 1.2-mm source number count density in MQN01 within $\abs{\Delta \varv_{\rm QSO}} < 1000\,{\rm km\,s^{-1}}$, and we compared with that of blank fields by selecting a sample of galaxies between $2.5 \le z \le 3.5$ from ASPECS large program \citep[see,][]{Aravena+2019, Aravena+2020, Boogaard+2019, Boogaard+2020, Gonzalez-Lopez+2019, Gonzalez-Lopez+2020}. This comparison yields to an overdensity signal which is also of the order of 10 -- 100 depending on the flux cut (see, Fig.~\ref{fig:12mm_LF}). We stress that such analysis is affected by large uncertainties due to low statistics of the selected samples. In addition, the selection of galaxies which are detected in their rest-frame sub-mm dust continuum emission and for which spectroscopic redshift information is available from different lines can introduce biases which are difficult to assess (see, Sect.~\ref{ssect:12mm_counts} for further details). Any interpretation here should therefore be taken with caution. Given the above limitations, an analysis of the source number count trend as function of continuum flux density cannot be performed. Notwithstanding, we argue that overdensity traced by dust continuum is consistent with what we found from the analysis of CO LFs. The source number counts shown of MQN01 in Fig.~\ref{fig:12mm_LF} do not exhibit any evidence of a flattening as observed in the CO LF, and can therefore be interpreted as a pure normalization effect with respect to that of blank fields. However, the 1.2-mm flux density at $z\sim3$ (corresponding to $\sim300\,{\rm \mu m}$ rest-frame wavelength) probes the RJ tail of the FIR dust emission which scales as $\sim M_{\rm dust}T_{\rm dust}$ \citep[see, e.g.,][]{Hodge+2021}. Therefore, any possible contribution to increasing the observed $1.2\,{\rm mm}$ flux density in galaxies located in the MQN01 volume cannot be directly attributed to either warmer dust temperature or a larger amount ($M_{\rm dust}$) of cold dust in galaxies. We also compare the observed 1.2-mm number counts with that toward the SSA22 protocluster at $z\sim 3$ \citep{Umehata+2017, Umehata+2019} obtained via a survey with similar design as we present here. The spectroscopic redshift information of galaxies in SSA22 are obtained through the measurement of CO(3--2) line thus providing a fair comparable sample. At zeroth order, the differential and cumulative number counts in MQN01 field are consistent with what has been found in SSA22 field. This evidence possibly suggests that similar physical mechanisms are caught in actions in these environments (we further discuss this point in Sect.~\ref{ssect:proto_comparison}).

\subsection{Molecular gas density in MQN01 field}\label{ssect:mol_dens}
We can estimate the total molecular gas content in the MQN01 structure by summing up the molecular gas masses of galaxies within the structure. To do so, we followed the recipe from (e.g.,) \citet{Decarli+2016, Jin+2021}
\eq{\rho({\rm H_{2}})\quadra{M_{\astrosun}{\rm \,cMpc^{-3}}} = \frac{1}{\Delta V}\sum_{i}M_{\rm H_{2}}^{i}\frac{F_{i}}{C_{i}},}
where $M_{\rm H_{2}}$ is the galaxy molecular gas mass. To estimate $M_{\rm H_{2}}$ for each galaxy, we have to assume a CO(4--3)-to-CO(1--0) conversion factor ($L'_{\rm CO(1-0)}=L'_{\rm CO(4-3)}/r_{41}$), and an $\alpha_{\rm CO}$ coefficient \citep[$M_{\rm H_{2}}=\alpha_{\rm CO}L'_{\rm CO(1-0)}$; see, ][for a review]{Bolatto+2013}. Such parameters are unknown for our CO(4--3) emitters in the MQN01 field. We adopted $r_{41}=0.61\pm0.13$ obtained via modeling of the CO ladders of sources at $z=2.0-2.7$ in the ASPECS field \citep[see,][and references therein]{Boogaard+2020}. The typical values of $\alpha_{\rm CO}$ adopted in the literature vary from $0.8\,M_{\astrosun}\,({\rm K\,km\,s^{-1}\,pc^{2}})^{-1}$ for local ULIRGs (ultra luminous infrared galaxies), SMGs, and quasar hosts \citep[][but see, also, \citealt{MotoyaArroyave+2023} who measured a higher average value of $1.7\pm0.5\,M_{\astrosun}\,{\rm (K\,km\,s^{-1}\,pc^{2})^{-1}}$ for local (U)LIRGs]{Downes+1998}, to $\sim4\,M_{\astrosun}\,({\rm K\,km\,s^{-1}\,pc^{2}})^{-1}$ for the giant molecular clouds (GMCs) in the Milky Way \citep[see,e.g.,][for further discussion]{Bolatto+2013, CarilliWalter2013}. To take into account all the uncertainties, we computed $\rho({\rm H_{2}})$ by performing a Monte Carlo Simulation by varying all the measurements within their uncertainties. For the source fidelity and completeness, we adopted the same approach as in the evaluation of the CO LFs, and continuum number counts (see, Sects.~\ref{ssect:co_lf} and~\ref{ssect:12mm_counts}). During the simulation, we also accounted for the uncertainty on $r_{41}$, and we varied $\alpha_{\rm CO}/(M_{\astrosun}\,({\rm K\,km\,s^{-1}\,pc^{2}})^{-1})$ uniformly in the range $0.8-4.3$. This procedure yields $\rho({\rm H_{2}})=3.7^{+1.0}_{-0.9}\times10^{8}\,M_{\astrosun}\,{\rm cMpc^{-3}}$, where the nominal values and uncertainties are computed by taking the 50th, 5th, and 95th percentile, respectively. Similarly, we computed the molecular gas density for galaxies within $\abs{\Delta\varv_{\rm QSO}}<1000\,{\rm km\,s^{-1}}$ ignoring completeness corrections and we obtained $\rho({\rm H_{2}})=(1.0\pm0.3)\times10^{9}\,M_{\astrosun}\,{\rm cMpc^{-3}}$.

   \begin{figure}[!t]
   	\centering
   	\resizebox{\hsize}{!}{
      		\includegraphics{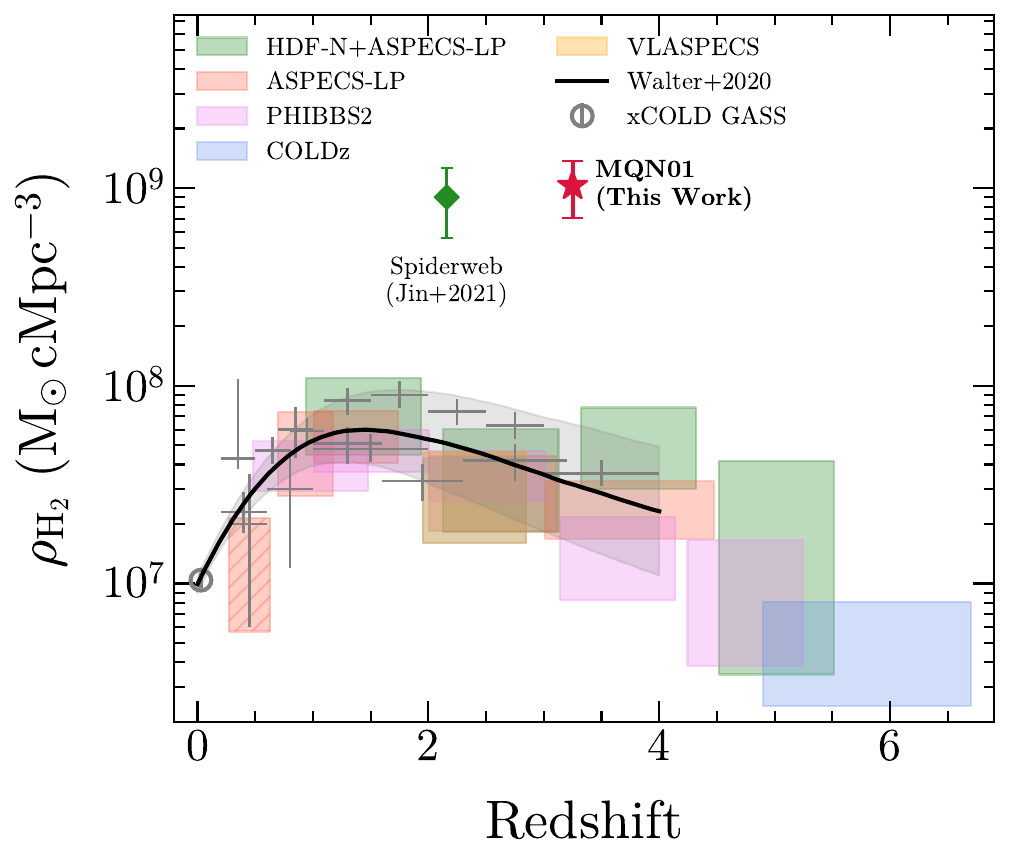}}
       \caption{Molecular gas density in the MQN01 field for galaxies within $\abs{\Delta\varv_{\rm QSO}}<1000\,{\rm km\,s^{-1}}$ (red star) compared to results in blank fields from HDF-N+ASPECS-LP \citep{Boogaard+2023}, ASPECS-LP \citep{Decarli+2019c, Decarli+2020}, PHIBBS2 \citep{Lenkic+2020}, COLDz \citep{Riechers+2019}, VLASPECS \citep{Riechers+2020b}, and the xCOLD GASS survey \citep{Fletcher+2021}. We also report measurements of the cosmic molecular gas density obtained via dust continuum \citep[gray bars;][]{Berta+2013, Scoville+2017, Magnelli+2020}, and the best-fitting function from \citet{Walter+2020}. The green symbol is the molecular gas density in the Spiderweb protocluster \citep{Jin+2021}.}
       \label{fig:rho_h2}
   \end{figure}

In Fig.~\ref{fig:rho_h2}, we compare our results with the molecular gas density across cosmic time in blank fields as reported by several works in the literature. We report results from the CO Luminosity Density at High Redshift \citep[COLDz;][]{Pavesi+2018,Riechers+2019}, ASPECS Large Program \citep{Decarli+2019c, Decarli+2020}, the Very Large Array (VLA) -- ASPECS \citep[VLASPECS;][]{Riechers+2020b}, the IRAM (Institute for Radio Astronomy in the Millimeter Range) Plateau de Bure High-$z$ Blue Sequence Survey 2 \citep[PHIBBS2;][]{Freundlich+2019, Lenkic+2020}, the NOEMA survey in the HDF-N \citep{Boogaard+2023}, and the xCOLD GASS survey \citep{Saintonge+2011, Saintonge+2017, Fletcher+2021}. We also report measurements of the cosmic molecular gas density obtained via the dust continuum survey \citep[see,][]{Berta+2013, Scoville+2017, Magnelli+2020}, as well as the best-fit to data obtained from \citet{Walter+2020}. We found that the molecular gas budget residing in MQN01 is $\sim 10\times$ the expected molecular gas density at $z\sim 3$ consistent with what we found from the analysis of the CO LF. This corroborates our findings indicating that galaxies in MQN01 are evolving in a gas-rich overdense environment. 

We also compare our result with that obtained in the Spiderweb protocluster at $z=2.16$ reported by \citet[][$\rho({\rm H_{2}})=9.0^{+3.6}_{-3.4}\times10^{8}\,M_{\astrosun}\,{\rm cMpc^{-3}}$]{Jin+2021} as traced by CO(1--0). The emission from this line is a more direct proxy (because it does not require assumptions on the CO line excitation) of the molecular gas mass in galaxies (albeit it still requires an assumption on the $\alpha_{\rm CO}$ luminosity-to-H$_{2}$ mass conversion factor). {\rm \citet{Jin+2021} adopted two typical values of $\alpha_{\rm CO}/(M_{\astrosun}\,({\rm K\,km\,s^{-1}\,pc^{2}})^{-1})$ for their sources equal to $0.8$ for starburst-like objects \citep{Emonts+2018} and $3.6$ for disk-like galaxies \citep{Daddi+2010}}. We note here that mm observations toward the Spiderweb protocluster probed a much larger volume ($\sim 6600\,{\rm cMpc^{3}}$) with respect to our ALMA survey in the MQN01 field . However, our measurements point to a molecular mass density which is comparable to that found in one of the most massive protoclusters at later times.

\subsection{Comparison with (sub-)mm surveys of high-$z$ protoclusters}\label{ssect:proto_comparison}
Comparing our findings with results obtained in other fields hosting high-$z$ galaxy-rich environments would allow us to put our results into a wider context. However, this is somehow difficult since a fair comparison requires at least a similar survey design at comparable redshift. Indeed, the observed galaxy population strongly depends on the observed tracers, selection method, sensitivity, and area of the survey. In our work we found a peculiar feature in the CO LF suggesting a flattening at the bright end relatively to the trend expected in blank fields. Despite numerous works in the literature reporting CO surveys in overdense environments (see, Sect.~\ref{sect:introduction}), a systematic analysis of the CO LF in such environments across the redshift range is still missing. \citet{Jin+2021} reported a large area ($\sim 21\,{\rm arcmin^{2}}$) survey of CO(1--0) with the Australia Telescope Compact Array (ATCA) toward the Spiderweb protocluster at $z=2.16$. By using 46 identified CO emitters, they put a first constraint of the CO(1--0) LF in a protocluster environment. By fitting the data using a Schechter function with a fixed slope as determined by studies in blank fields \citep[such as COLDz; see][]{Riechers+2019}, they found a CO(1--0) LF normalization which is $\sim 1.5\,{\rm dex}$ higher than that in fields. Interestingly, the CO(1--0) LF reported by \citet{Jin+2021}, does not seems to show any signature of flattening at its bright end, even when considering a small volume ($\sim 1650\,{\rm cMpc^{3}}$) centered around the starbursting radio galaxy MRC1138-262 \citep[see, e.g.][]{Miley+2006, Emonts+2016} where the highest concentration of galaxies has been found. 

Another attempt to study the LF in high-$z$ protocluster field is reported by \citet{Hill+2020}. The latter present ALMA follow-up observations of SPT2349-56, a star-forming protocluster core at $z=4.3$, targeting the CO(4--3) as well as the singly-ionized carbon fine-structure line \cii. They revealed 24 line-emitting sources in $\sim 7.2\,{\rm arcmin^{2}}$ through which they obtained the FIR and \cii{} luminosity differential number counts. Intriguingly, they found that the FIR LF in the core of SPT2349-56 is biased toward bright galaxies compared to what is predicted for field galaxies, thus suggesting an enhanced star-formation activity possibly triggered by ongoing mergers \citep[see, e.g.,][]{Tacconi+2008, Engel+2010, Luo+2015} which are expected to be more common in overdense regions \citep[see, e.g.,][]{Lotz+2013, Hine+2016}. On the basis of the almost-linear CO-FIR luminosity relation \citep[see, e.g.][]{ Liu+2015, Kamenetzky+2016} this would directly translate in an increased $L'_{\rm CO(4-3)}$ luminosity at the bright end of the CO LF as we observe in the MQN01 field.

A comparable example to our observations is the SSA22 field, a protocluster at $z\simeq3.1$ surveyed by \citet{Umehata+2015, Umehata+2017} with ALMA achieving a sensitivity of $\sim60\,{\rm \mu Jy\,beam^{-1}}$ over an area of $\sim2\arcmin\times3\arcmin$ probing the $1.1\,{\rm mm}$ continuum. Therefore this survey is fairly comparable to our work both in terms of survey design and redshift.  \citet{Umehata+2018, Umehata+2019} followed-up this field by extending both the probed area and including CO(3--2) line observations using ALMA band 3. In the combined survey, \citet{Umehata+2018} report the detection of 35 SMGs and the 1.1-mm number counts which reveal an excess by a factor of several with respect to blank field. In this work we combined data from \citet{Umehata+2018, Umehata+2019}, and we rescaled the observed fluxes to $1.2\,{\rm mm}$ at $z=3.25$ (see Sect.~\ref{ssect:12mm_counts}). The source number counts in MQN01 and SSA22 appear to be overlapped following a similar trend (albeit with large uncertainties). This might indicate that dust-obscured star-formation and metal production in galaxies in MQN01 field are enhanced in the protocluster. In addition, we conclude that the galaxy density as traced by the $1.2\,{\rm mm}$ dust continuum emission in MQN01 is comparable to that revealed in SSA22 field.

\subsection{Is MQN01 a protocluster in formation?}

\subsubsection{Galaxy velocity distribution}\label{ssect:vel_distr}
\begin{figure}[!t]
   	\centering
   	\resizebox{\hsize}{!}{
      		\includegraphics{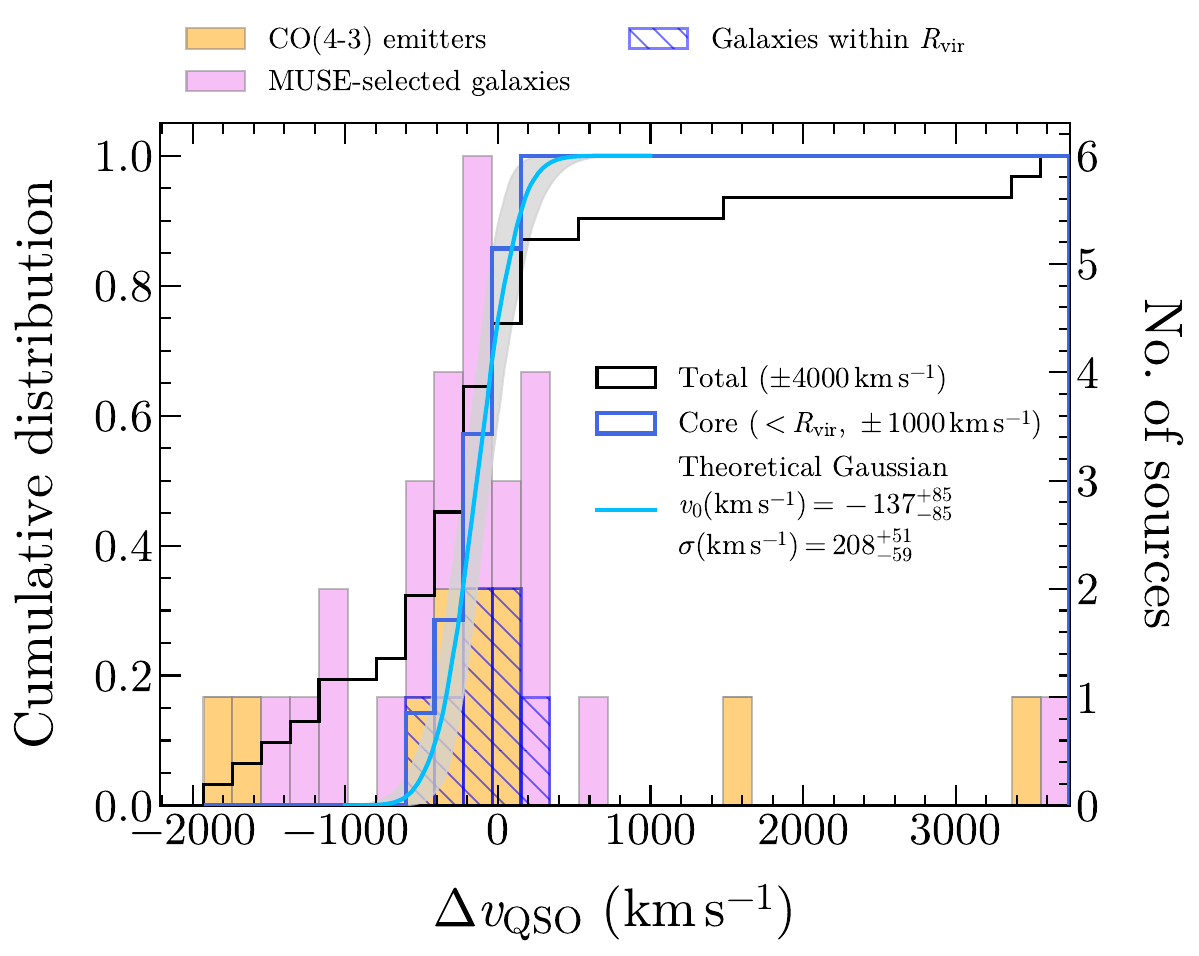}}
       \caption{{\it Right axis:} velocity distribution of galaxies in MQN01 field either detected in their CO(4--3) lines in our ALMA band 6 survey (orange bins), or having high-fidelity spectroscopic redshift within $\abs{\Delta\varv_{\rm QSO}}<4000\,{\rm km\,s^{-1}}$ measured from MUSE data (violet bins). The blue hatched bins are the distribution of galaxy velocities in the ``core''. {\it Left axis:} the black and blue lines are the cumulative distribution of the whole sample, and that of the ``core'', respectively. The light blue curve is the theoretical prediction for a Gaussian distribution of velocities with central value and dispersion as reported in the legend. The gray band represents the 1-$\sigma$ confidence interval.}
       \label{fig:v_distr}
   \end{figure}

In Fig.~\ref{fig:v_distr} we report the distribution of $\Delta\varv_{\rm QSO}$ of the combined ALMA+MUSE sample, including sources with high-fidelity spectroscopic redshift within $\abs{\Delta\varv_{\rm QSO}}< 4000\,{\rm km\,s^{-1}}$. Interestingly, we can identify a ``core'' population of galaxies having $\abs{\Delta\varv_{\rm QSO}}< 1000\,{\rm km\,s^{-1}}$ which appear to be Gaussian distributed as expected for cluster (i.e., virialized) galaxies \citep[see, e.g.,][]{Yahil+1977}, with the distribution median value appearing slightly shifted with respect to the QSO systemic redshift of $\sim-200\,{\rm km\,s^{-1}}$. To test this hypothesis, in Fig.~\ref{fig:v_distr} (blue bins) we additionally report the (cumulative) velocity distribution of galaxies belonging to $\abs{\Delta\varv_{\rm QSO}}< 1000\,{\rm km\,s^{-1}}$, {\rm within the estimated core virial radius of $R_{\rm vir}\sim 100\,{\rm kpc}$ (see, Sect.~\ref{ssect:halo_mass}).} %
We therefore tested the Gaussianity of the sample by performing a Lilliefors test \citep{Lilliefors1967}\footnote{This test is based on the commonly used Kolmogorov-Smirnov test. Lilliefors test is specifically used to test the null hypothesis that the sample comes from a normally distributed population, when the null hypothesis does not specify the expected value and variance of the normal distribution.} and we obtained a $p$-value of $0.94$. Therefore, we do not have significant evidence for rejecting the null hypothesis, but the small number statistics prevents here any strong conclusion. The mean velocity and the standard deviation of the ``core'' galaxies is $\varv_0 = -137^{+84}_{-86}\,{\rm km\,s^{-1}}$, and $\sigma=208^{+45}_{-59}\,{\rm km\,s^{-1}}$, where uncertainties are computed using bootstrap resampling. In Fig.~\ref{fig:v_distr} we report the theoretical cumulative Gaussian. The latter fairly reproduces the observed velocity distribution of the ``core'' galaxies.

\subsubsection{Halo mass estimate}\label{ssect:halo_mass}
Under the assumption that the observed galaxies in the ``core'' of MQN01 field are virialized within a large dark matter (DM) halo \citep[see, e.g.,][]{Miller+2018, Hill+2020}, we can estimate the total halo mass on the basis of the observed velocity dispersion of the galaxies. To do so, we employed the scaling relation found by \citet[][see, also \citealt{Rines+2013}]{Evrard+2008} which is based on a suite of $N$-body simulations with various cosmologies:
\eq{M_{\rm 200}=\frac{1}{h(z)}\tonda{\frac{\sigma_{r}}{1082.9\,{\rm km\,s^{-1}}}}^{1/0.3361}10^{15}\,M_{\astrosun},\label{eq:m200}}
where $h(z)=H(z)/100\,{\rm km\,s^{-1}\,Mpc^{-1}}$, and $\sigma_{r}$ is the line-of-sight velocity dispersion of the halo members. 

To define the halo members we proceeded iteratively. We took all galaxies within $\abs{\Delta\varv_{\rm QSO}}< 1000\,{\rm km\,s^{-1}}$, and we computed the dispersion of the line-of-sight velocity of galaxies (assuming they are due to purely peculiar motion), and the average position of the galaxies (i.e., the geometrical barycenter projected on the sky plane)\footnote{We note that the selected sample of galaxies does not change if we assume as reference velocity the mean redshift of the ``core'' sample instead of that of the QSO (see Sect.~\ref{ssect:vel_distr}).}. We therefore computed $M_{200}$ using Eq.~\ref{eq:m200}, and the virial radius ($R_{200}\equiv R_{\rm vir}$) as $R_{\rm vir}=[GM_{200}/(100\,H(z)^2)]^{1/3}$. We then iterated the procedure by selecting all galaxies within  $R_{\rm vir}$ until convergence. By doing so, we determined a total dynamical halo mass of $M_{200}=2.2^{+4.0}_{-1.8}\times10^{12}\,{M_{\astrosun}}$, and a virial radius of $R_{\rm vir}=94^{+37}_{-41}\,{\rm kpc}$, where the nominal values and uncertainties are computed with a Monte Carlo sampling and by taking the 5th, 50th, and 95th percentile of the distribution. However, we note that such estimates involve a set of assumptions and therefore has to be taken with caution. In particular, we have assumed that the halo is spherical and that thus the velocity dispersion can be simply derived from the line-of-sight velocity. However, numerical simulations show that halos are likely to be triaxial spheroids whose axis ratios depends on halo mass and that the line-of-sight velocity, on average, underestimate the dispersion but has significant scatter \citep[see, e.g.,][]{Elahi+2018b, Elahi+2018a}. Also, the kinematical analysis of cosmological simulations presented by \citet{deBeer+2023} suggest significant variations of the line-of-sight velocity of substructures within halos, due to asymmetries in matter accretion. Taken at face value, the mass estimate above would suggest that the host halo of the MQN01 quasar should be similar to the average quasar population at similar redshifts, (i.e., $M_{\rm 200}\simeq (1.5\pm0.5)\times10^{12}\,M_{\astrosun}$, see \citealt{deBeer+2023}, and references therein). This is somewhat in contrast with the large overdensity of CO emitters found around MQN01 on scales of a few cMpc, implying that similar overdensities should have been commonly found around the typical $z\sim3.5$ quasars, unless the MQN01 is a richer structure on larger scales still in the process of merging. In order to test this hypothesis, we estimated the clustering signal in our field as discussed in detail in Appendix~\ref{app:corr}. By fixing the slope of the cross-correlation function to $\gamma=1.8$ as commonly done in the literature at these redshifts \citep[see, e.g.,][]{Ouchi+2004, Diener+2017, Fossati+2021, Garcia-Vergara+2022}, we find a cross-correlation length for our CO emitters around the QSO CTS G18.01 of $r_{0,{\rm CO}}=16^{+13}_{-7}\,{\rm cMpc}$. However, we stress that the low number counts limits the statistical significance of our results in absolute terms. There are only very few other studies in the literature regarding the correlation length of CO emitters around typical quasars at $z>3$. By looking at the fields of 17 quasars at $z\sim3.8$ over an area of $\simeq66\arcsec$ radius per field, \citet{Garcia-Vergara+2022} built a sample of five CO emitters in a velocity window of $\pm 1000\,{\rm km\,s^{-1}}$ above the limiting luminosities (at $5.6\sigma$) that are dependent on distance from the quasars and that typically vary from $L'_{\rm CO}\simeq 4.5\times10^{9}\,{\rm K\,km\,s^{-1}\,pc^{2}}$ at the center of the field to $\simeq 3.1\times 10^{10}\,{\rm K\,km\,s^{-1}\,pc^{2}}$ at a distance of $\sim1.9\,{\rm cMpc}$ (rescaled following the primary-beam response of ALMA observations). We notice that our ALMA band 3 mosaic observation does not suffer from the same radial sensitivity dependence. In particular, \citet{Garcia-Vergara+2022} find a cross-correlation length of $r_{0,{\rm CO}}=12^{+4}_{-3}\,{\rm cMpc}$ (using the same fixed slope of $\gamma=1.8$) and a total overdensity of $\delta=17^{+12}_{-8}$. However, these estimates do not take into account a possible effect on the overdensity on the assumed luminosity cut. Our result shows that the overdensity increases with the CO emitter line luminosity (see, Fig.~\ref{fig:CO_LF}). Thus, we expect that a radial dependence of the sensitivity limits would artificially increase the overdensity as a function of radius and thus increasing the correlation length. We can directly test these expectations using our dataset. In particular, we have simulated in our data a similar sensitivity radial dependence as described above, obtaining a correlation length of $r_{0, \rm CO}=23^{+22}_{-10}\,{\rm cMpc}$ which is significantly larger with respect to the previous estimate with uniform sensitivity and with respect to the value found around typical quasars at $z\sim 3.8$ by \citet{Garcia-Vergara+2022} (albeit consistent within large uncertainties due to the low number statistics). This result suggests that either the MQN01 quasar host halo is much more massive than suggested by the kinematical analysis discussed above or that the quasar is surrounded by a rich structure consisting in multiple massive halos, or both. Ongoing surveys tracing the galaxy population on scales of tens of cMpc (Galbiati et al., in prep.) and aimed at characterizing the halos of other galaxies in the field will enable us to solve this dichotomy. 

  \begin{figure}[!t]
   	\centering
   	\resizebox{\hsize}{!}{
      		\includegraphics{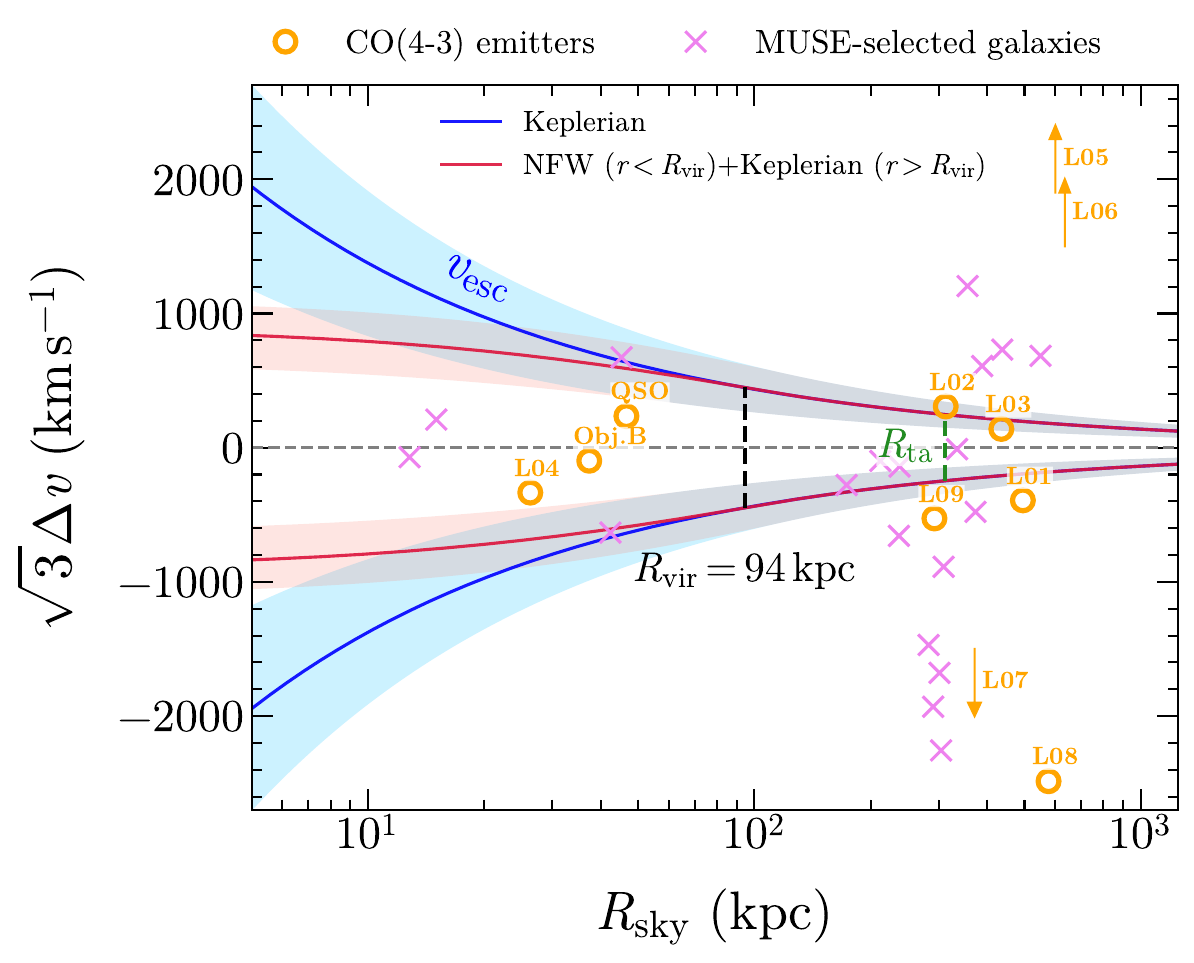}}
       \caption{Estimated three-dimensional galaxy velocities (including a statistical correction factor $\sqrt{3}$ assuming dynamical symmetry) as function of the projected separation with respect to the estimated halo center (also shown in Fig.~\ref{fig:selections}). The CO(4--3)-line emitting galaxies, and those with high-fidelity spectroscopic redshift from MUSE within $\abs{\Delta\varv_{\rm QSO}} <4000\,{\rm km\,s^{-1}}$ are indicated by orange circles, and violet crosses, respectively. The arrows indicate sources that lie outside the reported velocity range. The vertical dashed black and green lines mark the estimated virial ($R_{\rm vir}$) and ``turnaround'' radius ($R_{\rm ta}$). The blue and red curves are the predicted escape velocity corresponding to a central point-like object and a truncated NFW mass profile, respectively with a total mass equal to the estimated $M_{200}$. The shaded areas report the $1\sigma$ uncertainties. Galaxies which fall within the virial radius and the envelope are expected to be gravitationally bound.}
       \label{fig:v_esc}
   \end{figure}

In the assumptions that the halo mass is consistent with the kinematical analysis above, we can also estimate which galaxies are bound to the core by comparing their line-of-sight velocity as a function of the projected distance from the estimated center of the halo (i.e., the geometrical barycenter of galaxies within the ``core''; see, Fig.~\ref{fig:v_esc}). In particular, we shifted the galaxy velocities by their mean value within the virial radius and rescaled them by $\sqrt{3}$. We also report the escape velocity assuming a central point-like mass $\varv_{\rm esc}=\sqrt{2GM_{200}/R_{200}}$ as well as a more realistic model which is a truncated NFW mass profile \citep{Navarro+1996}\footnote{We truncated the NFW gravitational potential at $r=R_{\rm vir}$ and we kept it Keplerian outside. In this case $\Phi(r) = -(4\pi G\rho_{0}R_{s}^{3}/r) \ln(1+r/R_{s}) + K(c)G M_{200}/R_{\rm vir}$ for $r<R_{\rm vir}$, where $\rho_{0}=M_{200}/4\pi R_{s}^{3}f(c)$, $R_{s}=R_{\rm vir}/c$, $f(c)=\ln(1+c)-c/(1+c)$, $K(c)=c/[(1+c)\ln(1+c)-c]$, and we assumed a concentration index $c=4$ from \citet{Rodriguez-Puebla+2016}.}. Galaxies that fall within the envelope are expected to be gravitationally bound. However, sources which are located well beyond the virial radius are expected to be dragged by the Hubble flow, unless they have significant tangential velocity component toward the halo center. More precisely, galaxies that are expected to escape from the DM halo potential are those located beyond the so-called ``turnaround radius'' which is estimated from numerical simulations to be $R_{\rm ta}\simeq 3.3 R_{\rm vir}$ \citep[see, e.g.,][]{Gunn+1972, Busha+2003, Busha+2005}. This is the situation for a large fraction of galaxies, including 7 out of 11 CO-line emitters in our selected sample. Interestingly, we can identify a galaxy group (which includes sources L01, L02, L03, and L09) which lie within $\sim 200-500\,{\rm kpc}$ and $\abs{\Delta\varv}<1000\,{\rm km\,s^{-1}}$ from the inferred halo center. Such symmetrical configuration cannot be due to selection effect and it could suggest the existence of a causal connection between the ``core'', and a possible larger structure extending a few hundreds of kpc. In particular, L01 -- L03, and L09 show small line-of-sight velocity and are located close to the ``turnaround'' radius thus possibly suggesting that we are witnessing the moment in which these galaxies are decoupling from the Hubble flow and about to start their infall toward the core of the protocluster. However, we note that sky-projected distances of galaxies from the halo center has to be considered as lower limit to the true distances. Indeed, we expect to see in projection some galaxies having $R_{\rm sky}\sim R_{\rm ta}$ with three-dimensional distance $R > R_{\rm ta}$. These galaxies are physically in the Hubble flow and dynamically disconnected from the protocluster. This possibly explains why some galaxies (in particular several MUSE-selected sources), have a projected distance within (or comparable to) the ``turnaround'' radius while exhibiting high line-of-sight velocities consistent with the Hubble flow. Finally, we note that galaxies having large distance from the halo center also tend to show large velocity shifts. This is the case of L05 -- L08, which are all among the most distance (in the phase space) sources from the protocluster core and for which we do not have an independent redshift confirmation. Hence, in case future studies will determine that such sources are actually interlopers at different redshift or false-positive detections, this will not affect the dynamical analysis presented in this work. In conclusion, our analysis suggests that while the ``core'' appears to be bound, the entire system is not virialized. This conclusion is further supported by the measured galaxy overdensity considering that a virialized system should have a DM overdensity of around $\sim 200$ which is always smaller than that traced by galaxies \citep[see, e.g.,][]{Lacey+1994, Desjacques+2018}.

Finally, using the sources associated with the core, we can set a tentative limit of the baryon fraction in molecular form within the halo. We summed up the molecular mass of the individual CO-detected sources within the virial radius (i.e., QSO, Object B, and L04) which we estimated following the procedure described in Sect.~\ref{ssect:mol_dens}, assuming $\alpha_{\rm CO} = 0.8\,M_{\astrosun}\,({\rm K\,km\,s^{-1}\,pc^{2}})^{-1}$ typical of quasar hosts and SMGs \citep[e.g.,][]{Downes+1998}. By doing so, we found $M_{\rm H{2}}^{\rm core}=(8.8^{+0.9}_{-0.8})\times10^{10}\,M_{\astrosun}$ which is about $\sim 25\%$  of the total baryon mass of the halo assuming the maximum cosmic baryon fraction of $\Omega_{\rm b}/\Omega_{\rm m}=0.156$ \citep{PlanckColl+2020}. The molecular baryon fraction is therefore $f_{\rm mol}\equiv M_{\rm H{2}}^{\rm core}/ M_{200}< 0.04^{+0.07}_{-0.02}$ (assuming the estimated DM halo mass above as a lower limit). 

\section{Summary and Conclusions}\label{sect:summary_conclusions}
We presented an ALMA survey of the MQN01 field hosting a giant Ly-$\alpha$ nebula around a radio-quiet quasar at $z=3.25$ \citep{Borisova+2016}. Our survey is designed to primarily target the CO(4--3) transition as well as the 1.2-mm (rest-frame $\sim 300\,{\rm \mu m}$) dust continuum of galaxies in the entire MUSE FoV ($\sim 4\,{\rm arcmin^{2}}$). The combination of our FIR observations with the multiwavelength information collected using multiple facilities, allowed us to unveil the molecular gas content and the cold dust emission of galaxies in this field. Below, we summarize our main findings:

\begin{itemize}
\item We identified a robust sample of eleven CO(4--3) emitters within $\abs{\Delta\varv_{\rm QSO}}<4000\,{\rm km\,s^{-1}}$, including a closely-separated companion (Object B) $\sim 1\arcsec$ south to the QSO CTS G18.01. Object B was not detected previously in optical data due to its proximity to the QSO. Nine sources (including the QSO) are in the area covered by the MUSE observations, five of which ($\simeq 56\%$, including Object B) do not have any MUSE counterparts thus highlighting the crucial role of (sub-)mm surveys to obtain a complete census of galaxy population at high-$z$.
\item Our observations revealed eleven sources through their 1.2-mm dust continuum emission. Five of them (including the QSO, and Object B), are also detected in the CO(4--3) line, and another one having MUSE counterpart in the rest-frame UV. This implies the presence of six 1.2-mm-continuum selected galaxies in a narrow redshift range of $\abs{\Delta\varv_{\rm QSO}} < 1000\,{\rm km\,s^{-1}}$.
\item We analyzed the CO(4--3) LFs in MQN01 field within $2395\,{\rm cMpc^{3}}$ ($\abs{\Delta\varv_{\rm QSO}}<4000\,{\rm km\,s^{-1}}$), and $599\,{\rm cMpc^{3}}$ ($\abs{\Delta\varv_{\rm QSO}}<1000\,{\rm km\,s^{-1}}$) and we compare them with that of blank fields. We found a systematic excess of galaxy number counts which points to a galaxy overdensity of $\delta^{4000}_{\rm CO}(>L'_{\rm lim})=14^{+9}_{-6}$ and $\delta^{1000}_{\rm CO}(>L'_{\rm lim})=18^{+14}_{-8}$ above the limiting CO(4--3) luminosity of our ALMA survey. Notably, we found evidence of a flattening at the bright-end of the LF with respect to the expected trend in blank fields at similar redshift. Despite the limited statistics, this results suggest that massive galaxies in dense environments at $z\sim3$ are richer in molecular gas with respect to the field allowing them to grow faster than their counterparts in average environments.
\item We obtained the number counts density in MQN01 for sources detected in their $1.2\,{\rm mm}$ dust continuum which have high-fidelity spectroscopic information from rest-frame UV metal absorptions from MUSE, or CO line. By building a comparison sample for blank fields exploiting ASPECS catalogs, and applying proper ``$k$-corrections'', we found evidence for a systematic excess of source number counts. The overdensity revealed by our ALMA 1.2-mm survey is consistent (within the large uncertainties) with that traced by the molecular gas. Our source number count density is also consistent with that in the SSA22 protocluster at $z=3.1$ suggesting that obscured star-formation and consequently the metal production are similarly enhanced in such overdense environments. 
\item We studied the velocity distribution of galaxies in the field by using sources with high-fidelity spectroscopic redshift information from the MUSE and ALMA survey and we found evidence for a ``virialized'' structure which might represent the ``core'' of a protocluster in formation. Our dynamical analysis led to a total halo mass of $2.2^{+4.0}_{-1.8}\times10^{12}\,M_{\astrosun}$ which, however, should be considered as a lower limit on the true halo mass due to geometrical reason.
\end{itemize}

Further efforts are needed to deeply understand how galaxies are shaped during their evolution in dense environments and to shed light on the galaxy--large-scale structure connection. In particular, in future works we will fully characterize the galaxy population in MQN01 by exploiting our rich multiwavelength dataset, and we will compare the physical properties of galaxies (such as, e.g., their stellar and dust mass, SFRs, gas fraction, morphology, AGN contribution) with that of field galaxies at similar redshift. This will enable us to better assess the environmental effect on the galaxy assembly in one of the densest region of the Universe discovered so far at $z\sim3$. Through our high-resolution ALMA observations we will also study the ``core'' of the MQN01 structure by characterizing in details the galaxy morphology and kinematics, in particular those of the QSO host and the closely-separated Object B. In future studies, we will also explore the correlation between galaxy properties and the diffuse ionized gas on large-scale as probed by the Ly-$\alpha$ line emission. All these works will help us to dissect the galaxy assembly processes to address the question about how galaxies get their gas and how this affects the galaxy gas content.  

\begin{acknowledgements}
{\rm We thank the anonymous referee for the careful reading of the paper and useful suggestions which improved the manuscript.} This paper makes use of the following ALMA data: ADS/JAO.ALMA\#2021.1.00793.S. ALMA is a partnership of ESO (representing its member states), NSF (USA) and NINS (Japan), together with NRC (Canada), MOST and ASIAA (Taiwan), and KASI (Republic of Korea), in cooperation with the Republic of Chile. The Joint ALMA Observatory is operated by ESO, AUI/NRAO and NAOJ. This project was supported by the European Research Council (ERC) Consolidator Grant 864361 (CosmicWeb) and by Fondazione Cariplo grant no. 2020-0902. MM was supported in part by grant HST-GO-17065. This research made use of Astropy\footnote{\url{http://www.astropy.org}}, a community-developed core Python package for Astronomy \citep{AstropyI, AstropyII}, NumPy \citep{Numpy}, SciPy \citep{Scipy}, Matplotlib \citep{Matplotlib}, and Statsmodel \citep{Statsmodel}.

\end{acknowledgements}
%
%
%
\bibliographystyle{bibtex/aa}
\bibliography{bibtex/MyBib}

\begin{appendix} 
\section{Multiwavelength counterparts of ALMA-selected galaxies}\label{sect:mw_counterparts}\label{app:counterparts}
\begin{figure}[!htbp]
   	\centering
   	\resizebox{\hsize}{!}{
      		\includegraphics{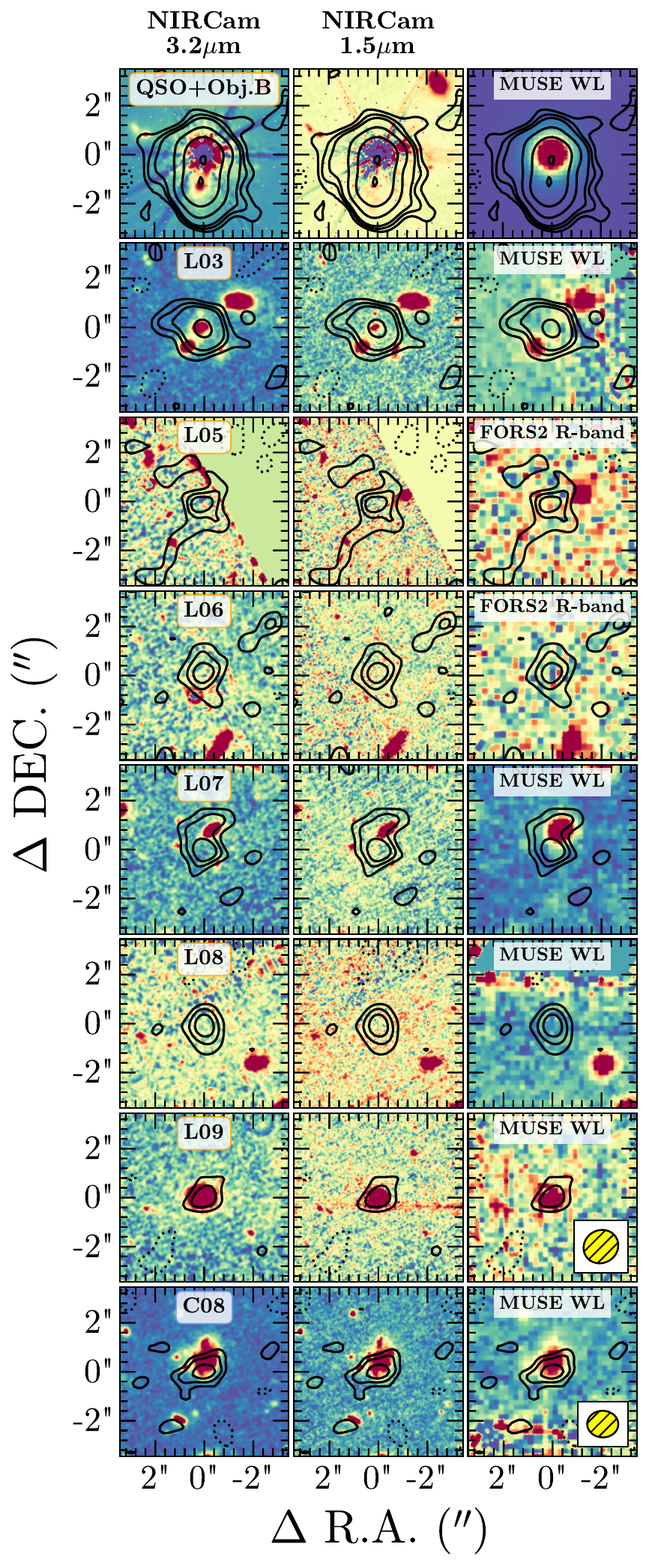}}
       \caption{Postage stamps ($7\arcsec\times7\arcsec$) of some ALMA-selected galaxies. {\it From the left to the right:} JWST/NIRCam F322W2 ($3.2\,{\rm \mu m}$), F150W2 ($1.5\,{\rm \mu m}$) and the MUSE white-light image. For sources which are located outside the MUSE FoV  we report instead the VLT/FORS2 R-band image. Black contours are superimposed showing the CO(4--3) line emission or, in the case of C08 the $1.2\,{\rm mm}$ dust continuum (bottom panels). Contours correspond to $[-2, 2, 3, 2^{N}]\times\sigma$ with $N>1$ integer number.}
       \label{fig:co_mw}
   \end{figure}

In Fig.~\ref{fig:co_mw} we report $7\arcsec\times7\arcsec$ postage stamps of galaxies with no rest-frame UV counterparts in MUSE (Object B, L03),  and/or remain undetected in the rest-frame optical in JWST/NIRCam images (i.e., L05, L06, L07, L08), as well as low-fidelity galaxies in our sample which have been selected by cross-matching the MUSE catalog of spectroscopically-confirmed objects (i.e., L09, C08). Such analysis capitalizes on our multiwavelength observational campaign (see, Sect.~\ref{ssect:ancillary_data}, for full details). In what follows, we report a qualitative discussion on the counterparts of such ALMA-selected galaxies studied in this work. 

The high luminosity of the QSO hinders both the optical/UV emission from its host and that of any other closely-separated galaxies. A preliminary subtraction of the QSO PSF in the JWST images reveals the south-east Object B in the FW322W2 filter and a possible extended component.

L03 is detected in ALMA CO(4--3) and the mm dust continuum (it corresponds to C04 selected with ALMA at 1.2 mm). While unambiguous counterpart is recognizable in the JWST/NIRCam images, especially in the long-wavelength filter, this object remains undetected in the rest-frame UV MUSE WL image. Notably, the closely-separated galaxy located at south-est is a high-confidence MUSE-selected LBGs within $\abs{\Delta\varv_{\rm QSO}}< 1000\,{\rm km\,s^{-1}}$. Therefore, these two objects might be located at small physical distance forming an interacting system.

The sources L05, L06, L07, and L08 are detected only in line in our ALMA band 3 mosaic. L05, and L06 lie outside the MUSE footprint, for these objects, in Fig.~\ref{fig:co_mw} we report the rest-frame UV VLT/FORS2 R-band image. As mentioned in Sect.~\ref{ssect:interlopers}, these sources might represent either actual false-positive detections or interlopers at different redshifts. We note that a relatively bright source is partially blended at north-west with the line emission of L07 observed in ALMA. The analysis of the MUSE spectrum unambiguously places the north-west source at $z\simeq0.3358$. At such redshift, no common astronomical-observed FIR transitions enters in the ALMA SPW in which we observe the emission line of L07. Therefore any possible association between the observed line emission in ALMA band 3 and the north-west source can be securely ruled out.

Finally, L09 and C08 have been selected via the cross-matching between low-fidelity ALMA-selected sources and high-fidelity MUSE-selected LBGs. In Fig.~\ref{fig:co_mw} we show their rest-frame optical/UV counterparts which are very well detected in both JWST/NIRCam and VLT/MUSE images.

\section{Clustering analysis}\label{app:corr}
We measure the QSO-CO emitters correlation length following \citet{Garcia-Vergara+2017, Garcia-Vergara+2022}. We expressed the volume-averaged projected cross-correlation function as
\eq{\chi(R) = \frac{1}{V_{\rm eff}}\int_{V{\rm eff}} \xi(R', Z)\,dV\label{eq:cross-corr}}
where $\xi(R, Z)$ is the three-dimensional cross-correlation function $\xi(r)=(r/r_{0})^{-\gamma}$ and $r_{0}$ is the correlation length. In Eq.~(\ref{eq:cross-corr}) we expressed the cross correlation function in terms of the QSO-galaxy distance components along the plane of the sky ($R$) and the line of sight ($Z$), which are related to the three-dimensional distance through $r^{2}=R^{2}+Z^{2}$. To estimate the correlation length in the MQN01 field, we considered all the sources within $\abs{\Delta\varv_{\rm QSO}}<1000\,{\rm km\,s^{-1}} =\Delta V_{1000}$ and we count the QSO-galaxy pairs in cylindrical shells centered on the QSO CTS G18.01. Each volume ($V_{\rm eff}$) has a depth along the line-of-sight of $\Delta Z \simeq (2\Delta V_{1000}H(z)^{-1})(1+z_{\rm QSO})$ (corresponding to $\simeq25.3\,{\rm cMpc}$) and a inner and outer transverse radius of $R_{\rm min}$, and $R_{\rm max}$, respectively. We measured $\chi(R)$ at angular scales $1\arcsec\le\theta\le75\arcsec$ (corresponding to $0.03 \le R\le 2.4\,{\rm cMpc}$ at $z\simeq 3.25$) in logarithmically-spaced bins through the following estimator:
\eq{\chi(R)=\frac{\left\langle {\rm QG}(R_{\rm min}\le R\le R_{\rm max})\right\rangle}{\left\langle {\rm QR}(R_{\rm min}\le R\le R_{\rm max})\right\rangle} - 1,}
where $\left\langle {\rm QG}(R_{\rm min}\le R\le R_{\rm max})\right\rangle$ and $\left\langle {\rm QR}(R_{\rm min}\le R\le R_{\rm max})\right\rangle$ are the number of QSO--CO-emitter pairs in our survey within the cylindrical shell volume and that expected in absence of clustering at similar redshifts, respectively.  The numbers of QSO-CO emitter pairs in MQN01 should then in principle be corrected for the completeness and fidelity of the sources. However, our sample of galaxies within $\Delta V_{1000}$ comprises six objects that are all secure sources with rest-frame optical/UV counterparts ($F=1$). Given the large uncertainties involved in this kind of analysis we then opted to do not apply the completeness correction and we assumed that each bin is dominated by Poisson noise for low count statistics \citep[see,][]{Gehrels1986}. To estimate the expected source number in blank fields we capitalized on the CO(4--3) LF at $z\simeq 3.8$ from ASPECS+HDF-N by \citet{Boogaard+2023} which is given by a Schechter function with parameters $\log \phi_{\ast}=-3.17^{+0.35}_{-0.38}$, $\log L_{\ast}=10.04^{+0.56}_{-0.24}$ at fixed slope of $\alpha=-0.2$. The number of expected pairs within a cylindrical shell volume is therefore given by
\eq{\left\langle {\rm QR}(R)\right\rangle = \Delta Z\int_{R_{\rm min}}^{R_{\rm max}}n_{\rm CO}(L'_{\rm CO}>L'_{\rm lim})\,2\pi R'\,dR',\label{eq:qr_pair}}
where $n_{\rm CO}(L'_{\rm CO}>L'_{\rm lim})$ is the number density of CO(4--3) emitters above a given limiting luminosity ($L'_{\rm lim}$). In our survey, the CO emitters within $\Delta V_{1000}$ are located in regions where our ALMA band 3 mosaic has uniform sensitivity. Therefore, in Eq.~(\ref{eq:qr_pair}) we take out $n_{\rm CO}(L'_{\rm CO}>L'_{\rm lim})$ from the integral assuming a constant limiting luminosity at $5\sigma$ of $L'_{\rm lim} =2.8\times10^{9}\,{\rm K\,km\,s^{-1}\,pc^{2}}$ for line width of ${\rm FWHM} = 331\,{\rm km\,s^{-1}}$ equal to the median line FWHM of the high-fidelity sample of CO(4--3) emitters from \citet{Gonzalez-Lopez+2019}. We note that in the computation of $\left\langle {\rm QR}(R)\right\rangle$ we also took into account the uncertainties on the CO(4--3) LF parameters. We note that this is crucial given that the LF is poorly constrained at its bright end thus introducing large uncertainties in the computation of the expected source number counts in blank fields. {\rm Specifically, we computed different realizations of $\left\langle {\rm QR}(R)\right\rangle$ for $30\times30$ couples of $(\log \phi_{\ast}, \log L_{\ast})$ extracted from random normal distributions with means and standard deviations corresponding to the nominal values and $1\sigma$ uncertainties reported by \citet{Boogaard+2023}, respectively. We then computed the uncertainty on $\left\langle {\rm QR}(R)\right\rangle$ by taking the 16th and 84th percentile of the distribution.}

 \begin{figure}[!t]
   	\centering
   	\resizebox{\hsize}{!}{
      		\includegraphics{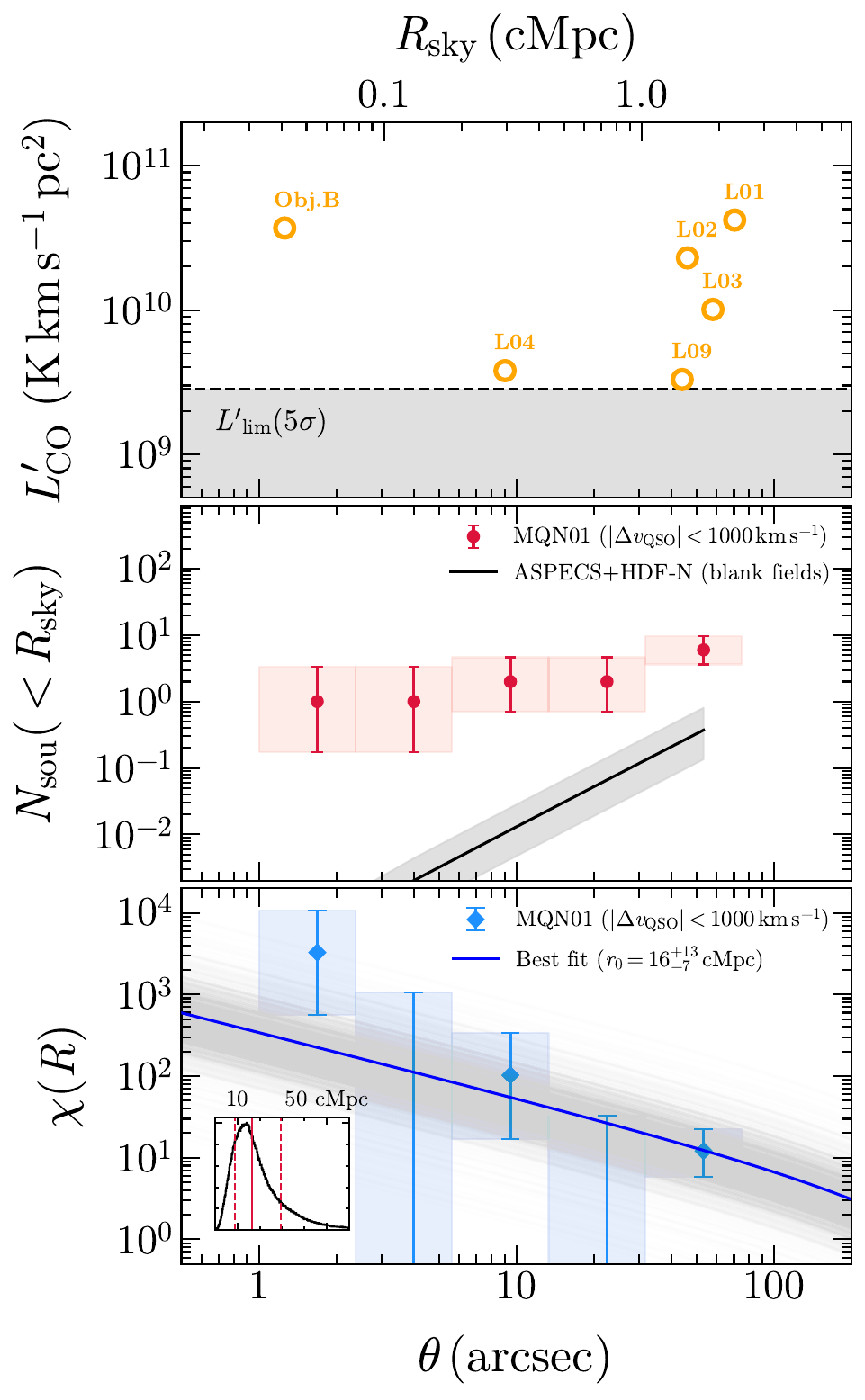}}
       \caption{Estimation of the cross-correlation length in MQN01. {\it Top panel:} source CO line luminosities as function of radial distance from the central quasar. The gray band shows the $5\sigma$ limiting luminosities of our survey (see text for further details). {\it Central panel:} Cumulative source number counts in MQN01 (red bins) and that expected in blank fields (black line, with their uncertainties shown in gray). {\it Bottom panel:} Projected volume-integrated cross correlation function in MQN01 (blue bins) and the best-fit model (blue solid line, and random draw from our Monte Carlo simulation in gray). The inset panel shows the posterior distribution of the three-dimensional cross-correlation length with red lines indicating the 16th, 50th, and 84th percentile, respectively.}
       \label{fig:corr_func}
   \end{figure}
   
To determine the three-dimensional cross-correlation length $r_{0}$, we fit our data by employing the Python MCMC ensemble sampler {\tt emcee} package \citep{Foreman+2013} adopting a Poisson maximum likelihood estimator \citep[see, e.g.,][]{Hennawi+2015, Garcia-Vergara+2017} and a fixed slope of $\gamma=1.8$ \citep[see, e.g.,][]{Ouchi+2004, Diener+2017, Fossati+2021, Garcia-Vergara+2017, Garcia-Vergara+2019,Garcia-Vergara+2022}. In order to properly take into account the uncertainties on the expected source number counts in blank fields, we repeated the fit 900 times by {\rm using the different realizations of $\left\langle {\rm QR}(R)\right\rangle$ computed as described above}. We then obtained the final posterior distribution of $r_{0}$ by summing-up all the distributions obtained with such procedure. We finally computed the best-fit $r_{0}$ value and its $1\sigma$ uncertainties by taking the 50th, 16th, and 84th percentile, respectively and obtained $r_{0, \rm CO}=16^{+13}_{-7}\,{\rm cMpc}$. We report our result in Fig.~\ref{fig:corr_func}.

There are very few other works in the literature that studied the clustering of CO emitters around quasars at $z>3$ \citep[e.g.,][]{Garcia-Vergara+2022}. Some of these works, make use of single-pointing ALMA observations to evaluate the QSO-galaxy cross correlation length using a small sample of CO emitters. In single-pointing observations, the primary beam response of ALMA antennas introduces a spatial dependence of the sensitivity (and therefore of the limiting luminosity), which therefore decreases radially from the center of the field toward the edge.

We tested the effect of a radial dependence of the limiting luminosity on the evaluation of the correlation length. We repeated the same procedure described above by rescaling our limiting luminosity at $5\sigma$ with the primary beam response of a individual pointing of our ALMA band 3 mosaic modeled as a 2D Gaussian with ${\rm FWHM}=53.6''$ (corresponding to the HPBW at the reference frequency of $108.5\,{\rm GHz}$; see, Table~\ref{tbl:obs_summary}).
This yields a higher (areal-averaged) limiting luminosity of $L'_{\rm lim} =3.3\times10^{10}\,{\rm K\,km\,s^{-1}\,pc^{2}}$ within the last radial annulus at $31.6\arcsec \le \theta\le 75\arcsec$. In this experiment, sources L03 and L09 would not be detected having $L'_{\rm CO}$ much below the $5\sigma$ limiting luminosity at their sky position and we therefore removed them from the MQN01 source number counts within $\Delta V_{1000}$ (see, Table~\ref{tbl:CO_cand}). In addition, the number of expected pairs in blank fields decreases in the last radial annulus due to the sharp drop of the CO(4--3) LF at such high luminosities ($L'_{\rm lim} > L_{\ast}$). The overall effect is an artificial boosting of the $\chi(R)$ value in the outer radial bin such that the fit of our data yields a larger best-fit correlation length of $r_{0, \rm CO}=23^{+22}_{-10}\,{\rm cMpc}$. This value is still consistent within the large uncertainties with the $r_{0, \rm CO}$ value previously found using the full sample of sources and a uniform limiting luminosities, but with a nominal value which is $\sim 40\%$ higher. This experiment shows how the variation of the limiting luminosity across the field of view might alter the estimated value of the cross-correlation length in case of low number statistics.



\end{appendix}

\end{document}